\documentclass[acmsmall]{acmart}

\usepackage{amsmath,amsfonts}
\usepackage{algorithmic}
\usepackage{graphicx}
\usepackage{textcomp}
\usepackage{xcolor}
\usepackage{caption}
\usepackage{comment}
\usepackage{xspace}
\usepackage{booktabs}
\usepackage{listings}
\usepackage{float}
\usepackage{longtable}
\usepackage{array}
\usepackage{url}
\usepackage[lined, algonl, ruled, boxed,vlined,linesnumbered]{algorithm2e} 
\usepackage{ragged2e}

\AtBeginDocument{%
  }

\newcolumntype{L}[1]{>{\raggedright\let\newline\\\arraybackslash\hspace{0pt}}m{#1}}
\newcolumntype{C}[1]{>{\centering\let\newline\\\arraybackslash\hspace{0pt}}m{#1}}
\newcolumntype{R}[1]{>{\raggedleft\let\newline\\\arraybackslash\hspace{0pt}}m{#1}}

\newcommand{\tool}{\textsc{MeCheck}\xspace}

\newcommand{\codefont}[1]{\footnotesize{\texttt{#1}}\normalsize}
\newcommand{\totalRule}{15\xspace}

\newcommand{\precision}{100\%\xspace}
\newcommand{\recall}{96\%\xspace}
\newcommand{\fscore}{98\%\xspace}
\newcommand{\totalProject}{831\xspace}

\newcommand{\totalReportedInReal}{{152}\xspace}
\newcommand{\totalInteresting}{{117}\xspace}
\newcommand{\totalInterestingEAs}{21\xspace}
\newcommand{\totalRealBugs}{{49}\xspace}
\newcommand{\totalFalsePositives}{35\xspace}
\newcommand{\totalInjectedProject}{45\xspace}
\newcommand{\totalInjection}{45\xspace}
\newcommand{\totalProjectIncluded}{115\xspace}
\newcommand{\totalRealProject}{70\xspace}

\setcopyright{cc}
\setcctype{by}
\acmDOI{10.1145/3715772}
\acmYear{2025}
\acmJournal{PACMSE}
\acmVolume{2}
\acmNumber{FSE}
\acmArticle{FSE053}
\acmMonth{7}
\received{2024-09-05}
\received[accepted]{2025-01-14}

\begin{document}

\title{Detecting Metadata-Related Bugs in Enterprise Applications}

\author{Md Mahir Asef Kabir}
\email{mdmahirasefk@vt.edu}
\orcid{0000-0001-6227-1816}
\affiliation{%
  \institution{Virginia Tech}
  \city{Blacksburg}
  \state{Virginia}
  \country{USA}
}

\author{Xiaoyin Wang}
\orcid{0000-0002-9079-5534}
\affiliation{%
  \institution{The University of Texas at San Antonio}
  \city{San Antonio}
  \state{Texas}
  \country{USA}
}
\email{xiaoyin.wang@utsa.edu}

\author{Na Meng}
\orcid{0000-0002-0230-5524}
\affiliation{%
  \institution{Virginia Tech}
  \city{Blacksburg}
  \state{Virginia}
  \country{USA}
}
\email{nm8247@vt.edu}

\begin{abstract}
When building enterprise applications (EAs) on Java frameworks (e.g., Spring), developers often configure application components via \emph{metadata} (i.e., Java annotations and XML files). It is  challenging for developers to correctly use metadata, because the usage rules can be complex and existing tools provide limited assistance.
When developers misuse metadata, EAs become misconfigured, which can trigger erroneous runtime behaviors or introduce security vulnerabilities. 
To help developers correctly use metadata, this paper presents (1) RSL---a domain-specific language that domain experts can adopt to prescribe metadata checking rules, and (2) \tool---a tool that takes in RSL rules and EAs to check for rule violations. 

With RSL, domain experts (e.g., owner developers of a Java framework) can specify metadata checking rules by defining content consistency among XML files, annotations, and Java code. Given such RSL rules and a program to scan, \tool interprets rules as cross-file static analyzers that scan Java and/or XML files to gather information and look for consistency violations.
For evaluation, we studied the Spring and JUnit documentation to manually define \totalRule rules, and created 2 datasets with \totalProjectIncluded open-source EAs.
The first dataset includes \totalInjectedProject EAs, and the ground truth of \totalInjection manually injected bugs. 
The second dataset includes multiple versions of \totalRealProject EAs. 
We observed that \tool identified bugs in the first dataset with \precision precision, \recall recall, and \fscore F-score. It reported \totalReportedInReal bugs in the second dataset, \totalRealBugs of which were already fixed by developers.
Our evaluation shows that \tool helps ensure the correct usage of metadata. 

\end{abstract}

\ccsdesc[500]{Software and its engineering~Specification languages}
\ccsdesc[500]{Software and its engineering~Interpreters}
\ccsdesc[300]{Software and its engineering~Software maintenance tools}
\ccsdesc[100]{Software and its engineering~Parsers}

\keywords{Domain-specific language, metadata, XML, annotation, consistency}

\maketitle

\section{Introduction}
Enterprise application (EA) development is a complex process of creating large-scale, multi-tiered, scalable, reliable, and secure network applications for business purposes~\cite{ea-development}. 
To reduce the complexity of EA development, developers usually build EAs on top of the Java EE platform~\cite{javaEE} or third-party frameworks like Spring~\cite{spring}. Such platforms or frameworks promote the principle of separation of concerns~\cite{Hursch95separationof}: 
they address nonfunctional concerns including persistence, transactions, and security, so that 
 developers only need to implement the core functionality of an EA by hand. 
Most of these platforms or frameworks support a declarative programming model, allowing EA developers to use metadata (i.e., Java annotations and XML files) when configuring (1) how application components are deployed, and (2) how these components interact with each other.

However, correctly using metadata can be challenging for developers due to three reasons. First, the usage rules are domain-specific and vary with Java frameworks, so developers can easily get confused and misuse metadata. Second, when a deployment or configuration task poses consistency constraints on the content of (i) Java annotations, (ii) code implementation, and/or (iii) XML files, developers may fail to observe all constraints when evolving software and thus inconsistently update metadata or code. Third, existing tools \cite{chamberlin2002xquery,Song12,Wen20} rarely analyze metadata together with Java code, let alone find semantic inconsistencies between software artifacts. Consequently, when metadata misuse leads to any misconfiguration \cite{misconfiguration}, abnormal runtime behavior \cite{springSecurity}, confusing error \cite{confuse_error}, or security vulnerability \cite{security-vul}, developers are on their own to handle metadata-related issues.

\begin{figure}
\centering
\includegraphics[width=.85\linewidth]{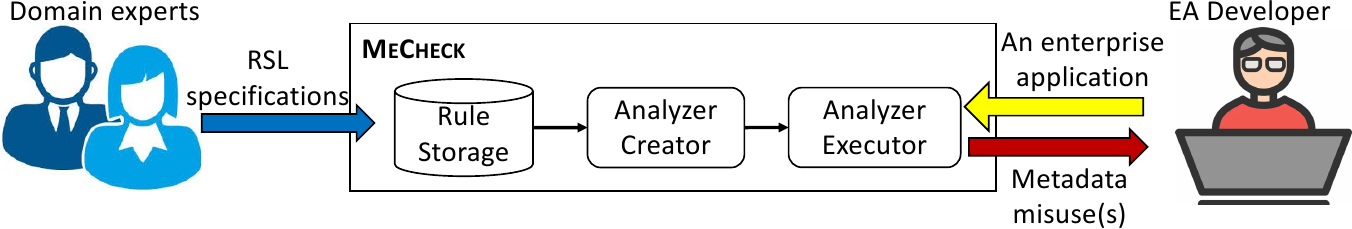}
\vspace{-.5em}
\caption{The workflow of our approach}\label{fig:overview}
\vspace{-1.em}
\end{figure}

To help developers debug metadata usage, we invented a semi-automatic solution that has two parts: \textbf{RSL (\underline{R}ule \underline{S}pecification \underline{L}anguage)} and \textbf{\tool (\underline{Me}tadata \underline{Check}er)}. As shown in Fig.~\ref{fig:overview}, to define a metadata-usage rule for a Java framework (e.g., Spring), domain experts (e.g., owner developers of that framework) can exploit RSL to (i) retrieve all relevant metadata and/or code items, (ii) refine those items based on certain conditions, (iii) constrain the content correspondence among refined items, and (iv) prescribe the error-reporting format if any constraint is violated. 
\tool stores such RSL specifications locally as domain-specific metadata checking rules. 
When an EA developer provides a software application to check, \tool loads all rules, 
creates parsing trees based on the RSL grammar for those rules, and treats the trees as analyzers. 
When executing each analyzer, \tool parses Java and XML files on demand, gathers and filters metadata and/or code items as instructed, and compares the item content for consistency checking. If any item violates a constraint, a customized error message is reported.

{To evaluate the effectiveness of RSL and \tool, we defined and investigated three research questions (RQs) in our experiments:
\begin{itemize}
\item \textbf{RQ1}: \emph{How effectively can RSL express metadata-usage rules?} We studied the documentation of Spring and JUnit frameworks~\cite{spring-doc,spring-tutorial,JUnit5}, and distilled \totalRule metadata-usage rules. 
Seven of the rules are about content consistency between \textbf{XML items (i.e., elements and attributes)} and Java code; six rules are about consistency checking between code and  annotations; one rule checks the consistency between XML and annotations; 
and one rule examines the consistency among code, XML items, as well as annotations. 
{We managed to express all rules using RSL, demonstrating its great power in expressing diverse rules.}
\item \textbf{RQ2}: \emph{How accurately can \tool detect bugs?} We manually injected \totalInjection metadata-related bugs into the latest version of \totalInjectedProject open-source projects, 
and applied \tool to those projects. Our evaluation shows that
\tool reported bugs with \precision precision, \recall recall, \fscore F-score. This implies that \tool detected bugs with high accuracy. 
\item \textbf{RQ3}: \emph{How effectively does \tool reveal real-world bugs?} We applied \tool to the version history of another \totalRealProject open-source projects, in order to reveal metadata-related bugs in real-world settings. 
{In total, \tool reported \totalReportedInReal bugs in the version history of \totalInterestingEAs projects, \totalInteresting of which bug reports are true positives.} 
Developers have fixed \totalRealBugs of those bugs so far. 
Our experiment indicates that \tool effectively identified real bugs in EAs. 
If developers had adopted \tool to scan their projects before committing any program changes, they could have found and addressed metadata-related issues more easily. 
\end{itemize}
}

In summary, this paper makes the following contributions: 
   	\begin{itemize}
   	    \item We designed a domain-specific language (DSL)---RSL---for domain experts to specify metadata-usage rules. Unlike prior DSLs, RSL can express consistency relations among XML items, Java annotations, and Java code. 
   	    \item We created \tool, to interpret RSL specifications and examine user-provided EAs accordingly. Compared with existing tools, \tool implements a novel algorithm that (1) extracts data in both Java and XML files, (2) tracks relations among the extracted data, and (3)  differentiates between data instances for refinement and comparison. 
   	    \item We comprehensively evaluated RSL and \tool. Our evaluation demonstrates the great expressiveness of RSL and the high detection accuracy of \tool for synthetic data; it also reveals real-world scenarios where \tool effectively locates metadata-related bugs. 
       \item To optimize \tool's runtime performance, we implemented a caching mechanism in the tool, which computes new data only when necessary.
   	\end{itemize}
    
In the following sections, we will  explain our research with a motivating example (Section \ref{se:example}), introduce the background knowledge of metadata usage (Section \ref{se:background}), describe our new approach: RSL and \tool (Sections \ref{se:rsl}--\ref{se:mecheck}), and present our evaluation (Section~\ref{se:evaluation}). 	
\vspace{-.5em}
\section{A Motivating Example}\label{se:example}

This section uses an example to intuitively explain our research. The software framework Spring~\cite{spring} supports developers to \emph{configure beans to have initialization and cleanup methods}~\cite{init-destroy}. 
Namely, a ``bean'' is 
any plain-old Java object that follows standard configuration patterns; it can be defined in Java or XML files.
Suppose that an XML file declares bean \codefont{b} as an instance of Java class \codefont{C}, and  
sets the bean's attribute ``\codefont{init-method}'' or ``\codefont{destroy-method}''  to a Java method name. Then the method must exist in the corresponding Java class \codefont{C}. This is because when the attribute is set, Spring automatically calls the corresponding method during runtime to either initialize 
\codefont{b} after the bean is created, or  perform destruction tasks before the bean is destroyed.

\begin{figure}[h]
\vspace{-.5em}
    \centering
    \includegraphics[width=\linewidth]{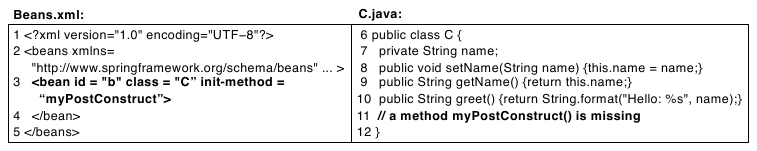}
    \vspace{-2.0em}
    \caption{{A bean in XML has \codefont{init-method} refer to a nonexistent method in the corresponding class}}
    \vspace{-.5em}
    \label{fig:example}
\end{figure}

As shown in Fig.~\ref{fig:example}, the exemplar XML file sets the attribute \codefont{init-method} of bean \codefont{b} to ``\codefont{myPostConstruct}'' (line 3). Ideally, EA developers should define a corresponding method in the exemplar Java class---\codefont{myPostConstruct()}. 
However, when EA developers fail to do so (see line 11), no existing compiler or static checker can reveal such omissions. 
As a result, developers can only observe the consequence of this metadata-related bug during program execution, and then manually diagnose root cause for the triggered runtime error \codefont{org.springframework.beans.factory.BeanCreationException}.

\begin{lstlisting}[label={lst:example-rsl},caption={An RSL specification {for detecting} missing methods}]
Rule method-exists {
  for (file xml in getXMLs()) {
    for (<bean> bean in getElms(xml, "<bean>")) {
      String beanClassFQN = getAttr(bean, "class");
      if (classExists(beanClassFQN)) {
        class c = locateClassFQN(beanClassFQN);
        for (String s in getAttrs(bean, "*method")) {
          assert(exists(method m in getMethods(c))(getName(m) == s)) {
            msg("The referenced method: %s in bean: %s is not defined in class: %s", s, getName(bean), getFQN(c));}}}}}} 
\end{lstlisting}

To help developers detect metadata-related bugs earlier and more easily, we developed RSL for framework developers to prescribe metadata-checking rules. Specifically, to statically detect the missing \codefont{init-method} mentioned above, framework developers can define an RSL rule. As shown in Listing~\ref{lst:example-rsl},  
intuitively, the rule describes four major things:  
\begin{itemize}
\item \textbf{What metadata/code items are involved?} Given a software project (i.e., EA), locate all $\langle bean\rangle$ objects defined in XML files (see lines 2--{3}).
\item \textbf{How are these items refined?} Associate the $\langle bean\rangle$ objects with their corresponding Java class declarations, and focus on such $\langle bean\rangle$-class pairs (lines {4}--6).
\item \textbf{What is the consistency constraint?} For attribute \codefont{<*method>} of any $\langle bean\rangle$ (i.e., \codefont{<init-method>} or \codefont{<destroy-method>}), a matching method should be 
defined in the related class (lines 7--8). 
\item \textbf{What is the error message?} When the constraint is violated and there is a <*method>-attribute value not matching any method definition, a bug should be reported (line 9). 
\end{itemize} 
Our new tool \tool can take in the above-mentioned RSL rule to statically check EAs. It can reveal the $\langle *method\rangle$  misconfiguration shown in Listing \ref{lst:example-rsl}, before EA developers run that program.
\vspace{-1.em}
\section{Background}\label{se:background}

To facilitate comprehension, 
this section will explain the three alternative ways of metadata-based EA configuration (Sections~\ref{ss:xml}--\ref{ss:xanda}), and our research problem
(Section~\ref{ss:problem}).

\vspace{-0.5em}
\subsection{XML-Based 
Configuration}\label{ss:xml}
There is a special kind of XML files named \textbf{deployment descriptors (DDs)}~\cite{DD_javaEE}, which are frequently used in EAs for configuration purposes. According to the XML syntax, 
\textbf{XML elements} are the basic building blocks of XML files~\cite{xml-elements}. 
Each element is used as a container to store text content, other XML elements, or attributes. The syntax of XML elements is:

\codefont{<element-name attributes> Contents ...</element-name>} 

\begin{figure}
    \centering
    \vspace{-1em}
    \includegraphics[width=\linewidth]{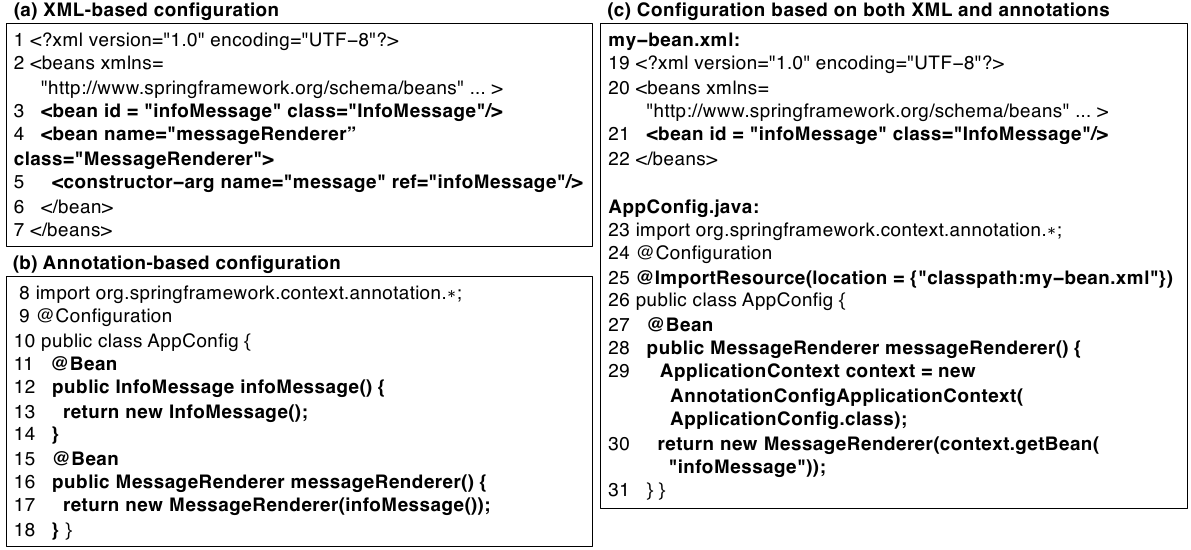}
    \vspace{-2em}
    \caption{{Three alternative ways to implement the same bean configuration}}
    \label{fig:examples}
    \vspace{-1.5em}
\end{figure}

\noindent 
The multiple attributes of any element are separated by white spaces, and each attribute associates the attribute name with a string value. For simplicity, we use \textbf{XML items} to refer to XML elements and their attributes.
For instance, Fig.~\ref{fig:examples} (a)  shows an exemplar DD that defines two bean objects: \codefont{infoMessage} and \codefont{messageRenderer} (see lines 3--6). The second bean has a \codefont{<constructor-arg>} element (line 5), which references the first bean as the value of its attribute \codefont{ref}. 

\vspace{-0.5em}
\subsection{Annotation-Based Configuration}\label{ss:a}
Recently, more and more Java frameworks provide annotations as a substitute for deployment descriptors~\cite{javaee-anno,spring-anno}. Annotations allow EA developers to specify component configurations within Java classes. 
An annotation can be attached to a Java class, method, or field. When annotations have attributes, the related syntax can be ``\codefont{@annotation-name("value")}'' or ``\codefont{@annotation-name(key = \{values\})}'', where \textbf{key} is the attribute name. 
 Fig.~\ref{fig:examples} (b) presents an annotation-based alternative to the DD shown in Fig.~\ref{fig:examples} (a). Specifically, \textbf{\codefont{@Configuration}} (line 9) implies that the current Java class is equivalent to a DD. Inside the class, \textbf{\codefont{@Bean}} (lines 11 and 15) is a direct analog of the XML \codefont{<bean>} element. \textbf{\codefont{@Bean}} 
means that the Java method it annotates will return an object that should be registered as a bean in the application context; the registered bean name is the method name (e.g., \codefont{infoMessage}). The second bean references the first bean by invoking \codefont{infoMessage()} (lines 15--18). 


\vspace{-0.5em}
\subsection{Combining XML Files with Annotations}\label{ss:xanda}

With both XML- and annotation- based configuration methods available, developers can take a hybrid approach between the two. Namely, given an EA, developers can configure part of the application with XML files and configure the remaining with annotations;  
the two types of configuration interact with each other in specialized ways. 
For instance, Fig.~\ref{fig:examples} (c) implements the XML-based configuration presented by Fig.~\ref{fig:examples} (a) with two files: {\sf my-bean.xml} defines the bean object \codefont{infoMessage} (line 21), and {\sf AppConfig.java} defines the other bean \codefont{messageRenderer} (lines 27--31). The Java class uses \codefont{@ImportResource} to import the XML file and to access all beans defined there. 

\vspace{-0.5em}
\subsection{Problem Statement}
\label{ss:problem}

As reflected by the examples so far, there are diverse metadata-usage rules that EA developers should follow. These rules can vary with the adopted software frameworks, developers' configuration needs, and configuration methods; many of the rules require for content consistency among code elements, XML items, and annotations.
 It is challenging for EA developers to correctly memorize and follow all domain-specific rules; it is even harder for developers to consistently maintain the metadata usage scattered in multiple files. A recent study on StackOverflow~\cite{meng2018secure} shows that numerous developers posted questions on how to properly configure frameworks.
There is a desperate need for approaches and tools that can help with metadata debugging. 
To create an automatic static checker that can satisfy developers' need, we encountered two technical challenges: (1) how to cope with the diversity of domain-specific rules, and (2) how to extract and relate the content from different software artifacts to validate metadata usage?

\vspace{-.5em}
\section{Rule Specification Language (RSL)}\label{se:rsl}

To overcome the challenges mentioned above, we designed a DSL---RSL---for framework developers to specify metadata-usage checking rules. The RSL rules will serve two purposes. First, they can help EA developers (i.e., framework users)  understand the content consistency among metadata and code items. Second, they will be sent to \tool for automatic detection of metadata-related bugs. 
As illustrated in Fig.~\ref{fig:syntax}, the RSL syntax defines statements, expressions, and built-in functions.

\vspace{-0.5em}
\subsection{Statements}
RSL supports five types of statements;  
each statement contains simple or complex expressions, or other statements. 
With these statement types, an RSL specification or rule describes four important aspects of the usage-checking logic:
extracting relevant metadata/code items, refining or filtering items, checking for consistency, and reporting errors.

(1) \textbf{ForStmt} means \codefont{for}-loop, used to describe \emph{what metadata/code items to extract and enumerate in a software application}. 
 Users can adopt a single or nested loop to structure the extraction and handling of all items potentially relevant to a constraint.  As shown in  Listing~\ref{lst:example-rsl}, \codefont{for(file xml in getXMLs())\{...\}} means to get all XML files in the project, iterate over those files using the variable \codefont{xml}, and process each iteration as instructed by the \codefont{for}-loop body. 

 (2) \textbf{IfStmt} is conditional \codefont{if}-statement. It describes \emph{how to refine items or locate items of interest}. Namely, when an iterated item satisfies the specified \codefont{if}-condition, it is processed by the body. 
For instance, 
in Listing~\ref{lst:example-rsl}, \codefont{if(classExists(beanClassFQN))\{...\}} checks whether a Java class with the same fully-qualified name as \codefont{beanClassFQN} exists in the source code. If so, 
the body first locates that class via \codefont{class c = locateClassFQN(beanClassFQN)}, and then
checks content correspondence between the Java class and XML file. Otherwise, the \codefont{bean}-object is skipped for further processing.  
This is because when the class is not defined in source code 
 (e.g., defined by a third-party library or by another languaged program), the content checking can become overly complicated or even infeasible; thus,  
 we decided to quit checking on the 
 \codefont{bean}-object to avoid false positives.

(3) \textbf{AssertStmt} means assertion, to describe \emph{what constraint to check}. Each statement has two parts: a condition and the body. 
The condition is a simple or complex expression, to define constraints or predicates that items must satisfy. The \codefont{assert}-body is defined with a MsgStmt (see below), to express action-to-take when the condition is not satisfied.
In Listing~\ref{lst:example-rsl}, the assertion means that given an attribute value of \codefont{<*method>}, there should be a Java method with the specified name. 


 (4) \textbf{MsgStmt} is message statement to describe \emph{what error message to report when a bug is detected}. People can use MsgStmt to compose an error-message template, and offer expressions or values to instantiate the template. 
 The MsgStmt in Listing~\ref{lst:example-rsl} incorporates a Java method name, bean name, and class name to pinpoint the issue of 
 a misconfigured \codefont{<init-method>} or \codefont{<destroy-method>}.

 (5) \textbf{DeclStmt} is declaration statement---an auxiliary statement to facilitate rule definition. Such a statement declares a variable with its data type, and initializes the variable using an expression. With such statements,
 recurring expressions only need to be evaluated once, as the evaluated value can be passed to a user-defined variable and that variable can get used to replace multiple occurrences of the same expression.
 For instance, Listing~\ref{lst:example-rsl} has a DeclStmt to declare variable \codefont{c} that 
 holds the located Java class item, given a \codefont{bean}-class name specified in XML. 

\begin{figure}[h]
\footnotesize
\vspace{-1em}
\begin{align*}
\textrm{Specification} := &\;\textbf{Rule}\ {\tt Id}\ \textrm{Body} \\
\textrm{Body} := &\;'\{' \textrm{Stmt}\ \textrm{Stmt}* '\}' \\
\textrm{Stmt} := &\;\textrm{ForStmt}\ |\ \textrm{IfStmt}\ |\ \textrm{AssertStmt}\ |\ \textrm{DeclStmt}\ ';' \\
\textrm{ForStmt} := &\;\textbf{for}\ '('\ \textrm{Type}\ {\tt Id}\ \textrm{in}\ \textrm{Exp}\ ')'\ \textrm{Body} \\
\textrm{IfStmt} := &\;\textbf{if}\ '('\ \textrm{Exp}\ ')'\ \textrm{Body}\\
\textrm{AssertStmt} := &\;\textbf{assert}\ '('\ \textrm{Exp}\ ')'\ '\{'\ \textrm{MsgStmt}\ ';'\ '\}' \\
\textrm{MsgStmt} := &\;\textbf{msg}\ '('\ ','\ \textrm{SimExp}\ (','\ \textrm{SimExp})*\ ')'\\
\textrm{DeclStmt} := &\;\textrm{Type}\ {\tt Id}\ '='\ \textrm{Exp}\\
\textrm{Exp} := &\;\textrm{SimExp}\ |\ \textrm{SimExp}\ \textbf{AND}\ \textrm{Exp}\ |\  \textrm{SimExp}\ \textbf{OR}\ \textrm{Exp}\ | \textbf{NOT}\ \textrm{Exp} \\
\textrm{SimExp} := &\;{\tt Id}\ |\ {\tt Lit}\ |\ \textrm{FunctionCall}\ |\ '('\ \textrm{Exp}\ ')'\ |\ \textrm{FunctionCall}\ '==' \textrm{SimExp}\ |\ \textbf{exists}\ '('\ {\tt Type}\ {\tt Id}\ \textbf{in}\ \textrm{Exp}\ ')'\ '('\ \textrm{Exp}\ ')' \\
{\tt Type} := &\;'\langle'\ {\tt Id}\ '\rangle'\ |\ \textbf{file}\ |\ \textbf{class}\ |\ \textbf{method}\ |\ \textbf{field}\ |\ \textbf{String} \\
{\tt Lit} := &\;{\tt StringLit}\ |\ {\tt CharLit}\ |\ {\tt IntLit}\ |\ {\tt FloatLit} \\
\textrm{FunctionCall} := &\;{\tt Id}\ '('\ \textrm{Params}\ ')' \\
\textrm{Params} := &\;\textrm{SimExp}\ (','\ \textrm{SimExp})* \\
\end{align*}
\vspace{-3.5em}
\caption{Core syntax of RSL}\label{fig:syntax}
\vspace{-2.em}
\end{figure}
 

\vspace{-0.5em}
\subsection{Expressions}
RSL defines various expressions. 
There are
 six kinds of simple expressions: identifier, literal, 
 invocation of built-in function(s), parenthesized expression, equivalence-checking for simple expressions, and \codefont{exists}-clause. In particular, the
\codefont{exists}-clause has two parts: a header and the body. The header \codefont{exists (Type Id in Exp)}
describes what items to enumerate, while the body \codefont{(Exp)} describes a constraint to satisfy by any item. This clause 
is similar to ForStmt, as it also describes \emph{what items to extract and enumerate}. However, different from ForStmt, this
clause does not have to process all items in a given set; it can exit early and return true whenever finding an item to satisfy the specified constraint.
In addition to simple expressions, RSL also supports three kinds of complex expressions, which connect simple expressions via logical operators AND, OR, and NOT.

\vspace{-0.5em}
\subsection{Built-in Functions}
As shown in Table~\ref{tab:built-in}, we defined four types of built-in functions to facilitate (1) data extraction from software artifacts and (2) content comparison among items.

\begin{table}
\caption{Built-in functions in RSL}
\footnotesize
\vspace{-1.4em}
\begin{tabular}{p{3.5cm}|p{9.8cm}}
\toprule
\textbf{Category} & \textbf{Functions} \\ \toprule
Code-related functions & \codefont{callExists, classExists, getArg, getClasses, getConstructors, getFamily, getFields, getFQN, getMethods, getName,  getReturnType, getSN, getType, hasField, hasParam, hasParamType, indexInBound,  isIterable, isLibraryClass, isUniqueSN, locateClassSN, locateClassFQN} \\\hline 
Annotation-related functions & \codefont{getAnnoAttr, getAnnoAttrNames, getAnnotated, hasAnnotation, hasAnnoAttr} \\ \hline 
XML-related functions & \codefont{elementExists, getAttr, getAttrs, getElms, getXMLs, hasAttr}\\ \hline 
Miscellaneous functions & \codefont{endsWith, isEmpty, indexOf, join, pathExists, substring, startsWith, upperCase} \\ 
\bottomrule
\end{tabular}
\vspace{-1.9em}
\label{tab:built-in}
\end{table}

(i) \textbf{Code-related functions} support data extraction from either raw Java code or processed code items (i.e., classes, fields, and methods). For instance, \codefont{classExists(String fqn)} checks whether the project defines a Java class, whose fully qualified name is \codefont{fqn}; 
\codefont{getClasses()} retrieves all classes defined by a Java project; 
\codefont{getFamily(class c)} retrieves all ancestors in the class hierarchy of a given class \codefont{c}, together with \codefont{c} itself; 
\codefont{hasParam(method m, String aName)} checks whether method \codefont{m} has a parameter named \codefont{aName}; 
\codefont{indexInBound(method m, int idx)} checks whether 
a Java method has at least \codefont{idx+1} parameters. 
Given a code element, \codefont{getName(...)}  returns the element's simple name, \codefont{getFQN(...)} returns the fully qualified name, and  \codefont{getType(...)} returns the type binding. 

(ii) \textbf{Annotation-related functions} extract information from annotations or annotated Java code. Specifically, \codefont{getAnnoAttr(class c, String anno, String attr)} returns the value of a specified attribute \codefont{attr}, of annotation \codefont{anno} used in Java class \codefont{c}; 
\codefont{getAnnoAttrNames(class c, String anno)} retrieves all attribute names of annotation \codefont{anno} in the given class \codefont{c}; 
\codefont{getAnnotated(String anno, String type)} retrieves all code items, which are decorated by the specified annotation and are of certain kind (e.g., Java class);
\codefont{hasAnnoAttr(class c, String anno, String attr)} checks whether the specified class has annotation \codefont{anno}, and whether that annotation has the attribute \codefont{attr}. 

(iii) The \textbf{XML-related functions} 
support data extraction from XML files, and the processing of XML elements or attributes.
 As shown in Listing~\ref{lst:example-rsl}, 
 \codefont{getXMLs()} returns all XML files in the project; 
 \codefont{elementExists(xml, "<bean>")} examines whether the given file has an XML element named as \codefont{<bean>}; 
 \codefont{getElms(xml, "<beans>")} returns all <bean>-elements in the given file;
 \codefont{getAttr(bean, "class")} returns the value of \codefont{class}-attribute for  \codefont{bean};
\codefont{getAttrs(bean, "*method")} retrieves values of \codefont{bean}'s attributes, whose names match the specified regular expression pattern. 


(iv) \textbf{Miscellaneous functions} define  operations applicable to Strings or Lists. For instance, \codefont{upperCase(String s)} capitalizes all letters in String \codefont{s};  
\codefont{join(List...values)} concatenates lists of items to create a larger list; 
\codefont{pathExists(String path)} checks whether a given path exists in the file system. 
\vspace{-.5em}
\section{\tool}\label{se:mecheck}
{\tool conducts {static program} analysis, to check whether metadata is used correctly. }
As shown in Fig.~\ref{fig:overview}, 
when applied to a Java enterprise application, 
\tool (i) loads RSL rules, (ii) creates analyzers from those rules, and (iii) executes them {to detect bugs.} This section focuses on steps (ii) and (iii) of this process. 

\vspace{-.5em}
\subsection{Analyzer Creator}
Given RSL specifications, this component produces parsing trees that we refer to as \textbf{analyzers}, because they reflect the logic of statically analyzing EAs to detect metadata-related bugs. We leveraged JavaCC \cite{javaCC_page} to implement the creator. 
Specifically, after we provided (1) token patterns defined in regular expressions and (2) syntax grammar defined in extended Backus-Naur form (EBNF), JavaCC derived a lexical analyzer (scanner) from the token patterns and created a parser from the grammar. 
The generated scanner and parser could create a parsing tree given an RSL rule. 

\vspace{-0.5em}
\subsection{Analyzer Executor}
When developing Analyzer Executor, we encountered three challenges (C1--C3):

\begin{itemize}
\item[C1.] \textbf{Semantics}: The five statement types supported by RSL {execute in different ways}, so our executor must observe the differences when interpreting their semantics.
\item[C2.] \textbf{Scoping}: If an RSL rule defines multiple variables using the same name, 
the executor should 
differentiate between the scopes of variables 
for correct static analysis. 
\item[C3.] \textbf{{Performance}}: When multiple rules require \tool to repetitively collect and process data in the same way (e.g., scanning all XML files to gather \codefont{<bean>}-elements), we need to reduce or even eliminate redundant computation to optimize performance.
\end{itemize}

To overcome all challenges, we created an
executor of analyzers (i.e., parsing trees) as 
a visitor to traverse tree nodes. 
When accessing each node, the executor collects and processes program data on demand; its traversal manner varies with the node type. 
For implementation, 
we used JavaParser \cite{javaparser} to parse Java code and extract data from the resulting parsing trees. We also used JDOM~\cite{oracle2014xmlparsers} to parse XML files and extract data. 
In this section, we will introduce our novel \emph{context-aware interpretation} algorithm to address C1 (Section~\ref{sec:alg}), the specially designed data structures to overcome C2 (Section~\ref{sec:data-structure}), and a novel caching mechanism to address C3 (Section~\ref{sec:cache}). 


\subsubsection{Context-Aware Interpretation Algorithm}\label{sec:alg}
As shown in Algorithm~\ref{alg:main}, 
given the parsing tree of analyzer $t$ and the software-under-analysis $P$, \tool implements a \emph{read-eval} loop to traverse statement-level nodes in a top-down manner, execute statements in sequence, and report
metadata-related bugs based on the logic reflected by $t$ (see lines 1.3--1.4). 
In particular, \tool adopts a stack $S$ to keep track of the \emph{execution context}, i.e., all variables defined and their scopes. During execution, \tool creates and pushes a frame $f$ before executing all statement-level child nodes under $t$ (line 1.2), 
but pops and destroys $f$ after executing those statements (line 1.5). 
Algorithms~\ref{alg:process_for}--\ref{alg:process_assert} reflect how \tool works differently when executing different kinds of statements.

\paragraph*{\textbf{ForStmt Execution.}}
As shown in Algorithm~\ref{alg:process_for}, \tool creates and pushes a frame $f$ (line 2.1), before executing a ForStmt's header and body in an iterative way. A typical way of defining the ForStmt header is: 
``\codefont{for (T v in func(...))}''.   
Given such a header, \tool calls \codefont{func(...)} to extract or gather data, and then uses a variable \codefont{v} of type \codefont{T} to enumerate data items. 
As shown in lines 2.5--2.9, in each loop iteration, 
\tool first adds one entry \codefont{\{T, v, e\}} to $f$, to associate \codefont{v} with its data type \codefont{T} and enumerated value \codefont{e} for that iteration; it then  executes statement-level child nodes in sequence; afterwards, it clears frame $f$ to remove all variables locally defined by or for that iteration (line 2.9), since the values of those variables are limited to that iteration.
After executing the entire \codefont{for}-loop, \tool removes $f$ to discard all variables locally declared (line 2.10). 

\paragraph*{\textbf{IfStmt Execution}}
As shown in lines 3.3--3.9 of Algorithm~\ref{alg:process}, 
\tool evaluates the \codefont{if-condition} expression, and proceeds to the statement's body when that expression is true.
Furthermore, before executing the body, \tool creates and pushes a frame $f$; it then executes statements inside the body sequentially; it also pops and discards $f$ after body execution.

\paragraph*{\textbf{The Execution of AssertStmt and MsgStmt}}
As shown in Algorithm~\ref{alg:process_assert}, \tool first evaluates the \codefont{assert}-condition. If that condition is  
false, \tool formulates a bug report based on the MsgStmt embedded in the \codefont{assert}-body. 

\begin{minipage}{.4\linewidth}
\footnotesize
\begin{algorithm}[H]
\caption{The \textbf{main()} function in our algorithm}\label{alg:main}
\KwIn{$t$---root node of the rule/analyzer, and $P$---the given software application}
\KwOut{Reported metadata-related bugs}
Initialize a stack $S$ for variable frames\\
Create a frame $f$ and push it onto $S$\\
\ForEach{statement node $c$ in $t$'s body}{
  $process(c, S, P)$
}
pop $f$ from $S$
\end{algorithm}
\begin{algorithm}[H]
\caption{The \textbf{processFor()} function}\label{alg:process_for}
\KwIn{$n$---statement node, $S$---stack for variable frames, and $P$---the given software application}
\KwOut{Reported metadata-related bugs}
Create a frame $f$ and push it onto $S$\\
$T \leftarrow$ variable type part from $n$'s header\\
$v \leftarrow$ variable name part from $n$'s header\\
$container \leftarrow evaluate($container part from $n$'s header$, S, P)$\\
\ForEach{element $e$ in $container$}{
  add \{$T$, $v$, $e$\} to $f$\\
  \ForEach{statement node $c$ in $n$'s body}{
    $process(c, S, P)$
  }
  clear the frame $f$
}
pop $f$ from $S$
\end{algorithm}
\end{minipage}
\hfill
\begin{minipage}{.52\linewidth}
\footnotesize
\begin{algorithm}[H]
\caption{The \textbf{process()} function}\label{alg:process}
\KwIn{$n$---statement node, $S$---stack for variable frames, and $P$---the given software application}
\KwOut{Reported metadata-related bugs}

  \If{$n$ is ForStmt}{
    $processFor(n, S, P)$
  }\ElseIf{$n$ is IfStmt}{
    $res \leftarrow evaluate(n$'s expression$, S, P)$\\
    \If{$res$ is True}{
        Create a frame $f$ and push it onto $S$\\
        \ForEach{statement node $c$ in $n$'s body}{
          $process(c, S, P)$
        }
        pop $f$ from $S$
    }
  }\ElseIf{$n$ is AssertStmt}{
      $processAssert(n, S, P)$
  }\ElseIf{$n$ is DeclStmt} {
      $f \leftarrow$ top of $S$\\
      $T \leftarrow$ variable type part of $n$\\
      $v \leftarrow$ variable name part of $n$\\
      $value \leftarrow evaluate($variable value part of $n, S, P)$\\
      add \{${T, v, value}$\} to $f$
  }
  \vspace{-.5em}
\end{algorithm}   
\begin{algorithm}[H]
\caption{The \textbf{processAssert()} function}\label{alg:process_assert}
\KwIn{$n$---statement node, $S$---stack for variable frames, and $P$---the given software application}
\KwOut{Reported metadata-related bugs}
$res \leftarrow$ $evaluate(n$'s expression$, S, P)$\\
\If{$res$ = False}{
  $msg \leftarrow$ $evaluate($message part of the assert statement, $S, P)$\\
  $print(msg)$
}
\end{algorithm}
\end{minipage}

\paragraph*{\textbf{DeclStmt Execution}} As shown in lines 3.12--3.17 of Algorithm~\ref{alg:process}, \tool retrieves the current top frame of stack $S$. It identifies both the type name \codefont{T} and variable name \codefont{v} used in the declaration; it also evaluates the right-hand side of the statement for \codefont{value}. Finally, \tool adds an entry \codefont{\{T, v, value\}} to $f$ to record the declared variable. In this way, when the variable is used later, \tool revisits $f$ to retrieve the variable's type as needed, and to read or write values on-demand.

\paragraph*{\textbf{Expression Evaluation}}
Algorithm~\ref{alg:evaluate} illustrates how expressions are processed and executed.
Intuitively, 
when an expression is a variable or literal, \tool simply returns the value as the evaluation result. When an expression is more complex and contains multiple sub-expressions or nested expressions, \tool traverses the expression tree in a top-down manner, and evaluates values in a bottom-up manner. For instance, to evaluate the expression \codefont{locateClassFQN(getAttr(bean, "class")}, \tool first retrieves the value of \codefont{bean} by accessing the stack; it then evaluates the value of function call \codefont{getAttr(bean, "class")}; based on that evaluation, \tool further calculates the value of function call \codefont{locateClassFQN(...)}.

\vspace{-.5em}
\subsubsection{Stack and Frames}
\label{sec:data-structure}
\tool adopts a stack $S$ to keep track of the execution context; it (1) pushes frames onto $S$, (2) pops frames from $S$, and (3) walks through frames to manage variables, their types, as well as their values. A \textbf{frame} is a dictionary, where the key is a variable name, and the value is a pair of data type and variable value.
\tool creates a frame before executing the entire analyzer, any ForStmt, IfStmt, or \codefont{exists}-clause; each created frame is then pushed onto stack to record any variable declared for or in the corresponding program structure. During execution, 
\tool refers to the stack to search for declared variables, query their data types, or obtain the values. 
After the execution of a ForStmt, IfStmt, exists-clause, or the entire analyzer, \tool pops a frame from the stack to discard all variables created by that program structure. 


\vspace{.2em}
\begin{minipage}{0.95\linewidth}
\begin{algorithm}[H]
\footnotesize
\caption{The \textbf{evaluate()} function}\label{alg:evaluate}
\KwIn{$n$---expression node, $S$---stack for variable frames, and $P$---the given software application}
\KwOut{Return a boolean value based on the expression evaluation}

\If{$n$ is an identifier}{
    $res \leftarrow$ retrieve the value of $identifier$ from $S$ starting from top frame\\
    $return$ $res$
  }\ElseIf{$n$ is a literal} {
   $return$ the literal's value
  }\ElseIf{$n$ is a built-in function call} {
    $return$ the function call's return value
  }\ElseIf{$n$ is a parenthesized expression} {
    $return$ $evaluate($the expression inside the parenthesis$, S, P)$ 
  }\ElseIf{$n$ is an equivalence-checking expression}{
    $return$ $evaluate($left-side of the expression$, S, P)$ == $evaluate($right-side of the expression$, S, P)$
  }
  \ElseIf{$n$ startsWith "exists"}{
  Create a frame $f$ and push it onto $S$\\
  $T \leftarrow$ variable type part from $n$\\
  $v \leftarrow$ variable name part from $n$\\
  $container \leftarrow evaluate($container part from $n, S, P)$\\
  \ForEach{element $e$ in $container$}{
    add \{$T$, $v$, $e$\} to $f$\\
    $logic \leftarrow$ expression inside the exists-expression's body\\
    $res \leftarrow$ $evaluate(logic, S, P)$\\ 
    \If{$res$ = True}{
      $return$ True
    }
    clear the frame $f$
  }
  pop $f$ from $S$\\
  $return$ False
}\ElseIf{$n$ is an AndExpression}{
    $return$ $evaluate($left-side of the expression$, S, P)$ \&\& $evaluate($right-side of the expression$, S, P)$
}\ElseIf{$n$ is an OrExpression}{
    $return$ $evaluate($left-side of the expression$, S, P)$ || $evaluate($right-side of the expression$, S, P)$
}\ElseIf{$n$ is a NotExpression}{
    $return$ ~$evaluate($the expression after NOT operator$, S, P)$
}
\end{algorithm}
\end{minipage}
\vspace{.2em}

\begin{figure}
\centering
\includegraphics[width=.88\linewidth]{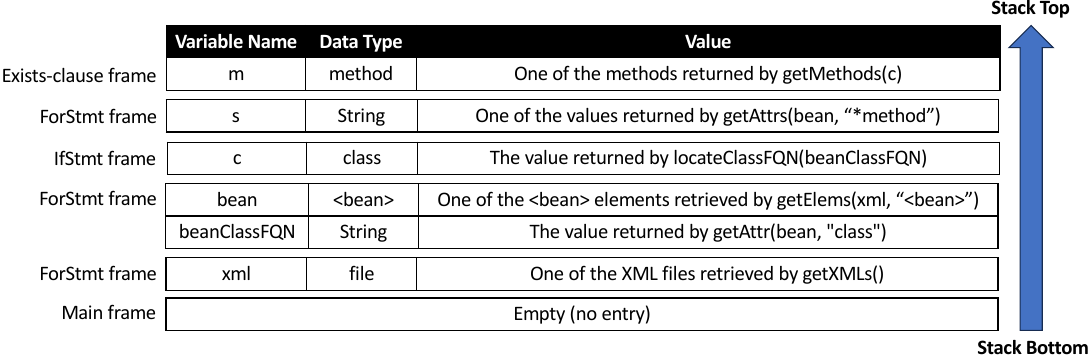}
\vspace{-.5em}
\caption{The stack status when \tool executes the {\tt exists}-clause in Listing~\ref{lst:example-rsl}}\label{fig:data-structure}
\end{figure}

Fig.~\ref{fig:data-structure} presents a snapshot of stack $S$ during runtime  when \tool executes the \codefont{exists}-clause of analyzer in Listing~\ref{lst:example-rsl}. Variable entries are added to frames when the variables are declared, or when iterating variables for ForStmt and \codefont{exists}-clause are initialized at the beginning of each loop iteration. Some entries get removed from frames at the end of loop iterations, because the values of iterating variables get reset for each iteration and the local variables declared for one iteration are invalid for other iterations. 
Given a variable name for resolution, e.g., \codefont{c} used in the \codefont{exists}-body, \tool starts with the top frame to search for a corresponding variable declaration. As shown in Fig.~\ref{fig:data-structure}, the top frame has no variable declared as \codefont{c}, so \tool moves on to the next frame---the {ForStmt} frame---to search. Such a stack walking continues until \tool finds the variable declared in a {IfStmt} frame. 
Our stack-based data structure allows the analysis-logic implemented in inner scopes to freely access variables defined in outer scopes; it also enables \tool to differentiate between the same-named variables declared in distinct scopes. 

  \vspace{-.5em}
\subsubsection{Performance Improvement}
\label{sec:cache}

When \tool executes multiple analyzers to scan the same project in one run, it is possible that the analyzers call same functions and repetitively retrieve the same results. 
Such repetitive function calls can waste lots of computing resources, hindering \tool from analyzing software efficiently. 
What makes things even worse, based on our experience, many domain-specific constraints or RSL analyzers
require \tool to parse all Java classes by calling \codefont{getClasses()}, or parse all XML files by calling \codefont{getXMLs()}. When a software project is large and contains thousands of files, the repetitive file-parsing can be very time-consuming. 

\begin{figure}
    \centering
    \includegraphics[width=.95\linewidth]{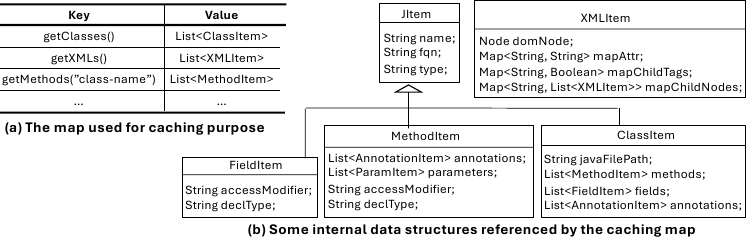}
      \vspace{-.5em}
    \caption{The map used for caching and some related internal data structures in our implementation}
    \label{fig:cache}  
\end{figure}

To optimize performance, we implemented a caching mechanism in \tool to eliminate redundant computation as much as possible. Specifically, we defined a global cache to map function calls to their return-values, for each software-under-analysis. For instance, as shown in Fig.~\ref{fig:cache} (a), \codefont{getClasses()} is mapped to the list of class items \tool creates based on Java parsing; \codefont{getXMLs()} is mapped to a list of XML items \tool creates based on XML parsing; \codefont{getMethods("class-name")} is mapped to a list of method items created for all methods declared by the Java class named ``\codefont{class-name}''. 
We decided to cache a function call if that function extracts data from Java/XML files, or derives information by enumerating extracted data. With such a caching mechanism, for each project-under-analysis, \tool parses all Java (or XML) files only once if any of the analyzers needs to retrieve all code (or metadata) items; when a function is called with different parameter values, their results can be cached differently as those values are used as part of the cache key.
Furthermore, to avoid eagerly extracting all methods and fields from Java files when \tool is only interested in the class-level information, \tool conducts lazy initialization for both \codefont{fields} and \codefont{methods} members of \codefont{ClassItem} objects (see Fig.~\ref{fig:cache} (b)): 
either member is initialized only when any analyzer explicitly requests for the information of a particular class.

\vspace{-.5em}
\begin{table}[h]
\centering
\footnotesize
\caption{The \totalRule metadata-usage constraints we summarized}\label{tab:constraints}
  \vspace{-1em}
\begin{tabular}{|c|p{1.5cm}|p{8cm}|p{1.5cm}|}
\hline
\textbf{Id} & \textbf{Rule Name} & \textbf{Constraint Summary} & \textbf{Involved Items} \\
\hline
$r_1$ & xml-path-check & When constructor ``{\tt ClassPathXMLApplicationContext(String configLocation)}'' is called, the provided argument should correspond to an existent XML file in the file system. &XML, code \\
\hline
$r_2$ & bean-class-exists & Java classes mentioned by the {\tt class}-attribute of {\tt <bean>} should exist in the project, unless they are library classes. & XML, code \\
\hline
$r_3$ & constructor-arg-type-field-map & When {\tt <constructor-arg>} is specified as a sub-element of {\tt <bean>}, its {\tt type}-attribute value should match a {constructor parameter}'s type name of the corresponding Java class.&XML, code \\
\hline
$r_4$ & constructor-arg-name-field-map & The {\tt name}-attribute value of {\tt <constructor-arg>} should match a {constructor parameter} name in the corresponding Java class.& XML, code \\
\hline
$r_5$ & constructor-index-out-of-bound & The {\tt index}-attribute value of {<constructor-arg>} should fit into the index boundary of at least one constructor in the corresponding Java file. & XML, code \\
\hline
$r_6$ & method-exists & When a {\tt <bean>} has any of the following attributes configured: {\tt init-method} and {\tt destroy-method}, the attribute values should match the names of methods defined in corresponding Java classes. &XML, code \\
\hline
$r_7$ & property-setter-map & The {\tt name}-attribute values of {\tt <property>} items should map to setter methods' names in the corresponding Java class. For instance, if the {\tt name}-attribute of a {\tt <property>} item is ``{\tt pn}'', then there should be a setter named ``{\tt setPn}'' in the corresponding Java class. & XML, code \\
\hline
$r_{8}$ & runwith-no-parameters & If a Java class has ``{\tt @RunWith(Parameterized.class)}'' annotation, then it should also have a method with ``{\tt @Parameters}'' annotation. & annotation, code\\
\hline
$r_{9}$ & runwith-no-test & If a Java class has ``{\tt @RunWith(Parameterized.class)}'' annotation, then it should also have a method with ``{\tt @Test}'' annotation. & annotation, code\\
\hline
$r_{10}$ & runwith-no-suiteclasses & If a Java class has ``{\tt @RunWith(Suite.class)}'' annotation, then the class should also have ``{  \tt @SuiteClasses}'' annotation. & annotation, code \\
\hline
$r_{11}$ & suiteclasses-no-runwith & If a Java class has ``{\tt @SuiteClasses}'' annotation, then the class should also have ``{\tt @RunWith(Suite.class)}'' annotation. & annotation, code \\
\hline
$r_{12}$ & suiteclasses-no-test & If a Java class is annotated with ``{\tt @SuiteClasses}'' or ``{\tt @Suite.SuiteClasses}'', then the Java class mentioned in the annotation attribute should (1) be decorated with either annotation, or (2) have a method in itself or any ancestor class satisfying any of the following requirement: (i) the method is annotated with {\tt @Test}, (ii) the method name starts with ``{\tt test}'', and (iii) the method name is ``{\tt suite}''. & annotation, code
\\
\hline
$r_{13}$ & testParams-not-iterable & Each Java method with ``{\tt  @Parameters}'' annotation should have an iterable return type (e.g., List). & annotation, code \\
\hline

$r_{14}$ & import-resource-path & When ``{\tt @ImportResource}'' is used and its attribute ``{\tt location}'' is configured, the attribute value should be a valid file path corresponding to an existent XML file in the file system. & XML, annotation \\
\hline

$r_{15}$ & bean-exists & The Spring API ``{\tt ApplicationContext.getBean(String str)}'' searches for a bean by name or by type. 
Search-by-type works when the argument {\tt str} is a Java class name (i.e., ending with ``{\tt .class}''), and the class should be either (1) annotated with ``{\tt @Component}'', ``{\tt @Service}'', ``{\tt @Repository}'', ``{\tt @Controller}'', or ``{\tt @RestController}'', or (2) mentioned by the {\tt class}-attribute of any {\tt <bean>} in XML. 
Search-by-name requires either a Java method annotated with ``{\tt @Bean}'' is named with {\tt str}, or a {\tt <bean>} in XML has its {\tt  id}-attribute value as {\tt str}.& XML, annotation, code \\
\hline
\end{tabular}
\vspace{-2.5em}
\end{table}

\section{Evaluation}\label{se:evaluation}

To assess the usefulness of our approach, we conducted three experiments and explored the following research questions (RQs):

\begin{itemize}
\item \textbf{RQ1}: How effectively can RSL express metadata-usage rules?
\item \textbf{RQ2}: How accurately can \tool detect bugs?
\item \textbf{RQ3}: How effectively does \tool reveal real-world bugs?
\end{itemize}

  \vspace{-.5em}
\subsection{Effectiveness of RSL}\label{ss:eval-effect-rsl}

To assess the expression power of RSL, we studied (1) the documentation of Spring \cite{spring}---the most popular application development framework for EAs in Java, and (2) documentation as well as tutorials on JUnit~\cite{junit-doc,junit-tutorial,how-to-junit,software-testing}---the most popular framework for unit testing in Java. 
We manually distilled \totalRule metadata-usage constraints from those documents. As shown in Table \ref{tab:constraints}, seven constraints are about the correspondence between XML and code ($r_1$--$r_7$); six constraints involve annotations and code ($r8$--$r13$); one constraint is on the correspondence between XML and annotation ($r14$); one constraint is relevant to XML, annotation, and code.
These constraints are diverse in terms of the metadata/code items involved, the number of files scanned, and the consistency-checking logic. We successfully expressed constraint-checking rules in RSL, which fact  evidences the great expression power of our domain-specific language. In the following experiments, we will leverage \tool to generate analyzers from these rules, and evaluate \tool's effectiveness in identifying buggy EAs that violate any of the rules.

\vspace{0.5em}
\noindent\begin{tabular}{|p{13.6cm}|}
	\hline
	\textbf{Finding 1 (Answer to RQ1):} \emph{RSL has great expressiveness. It is capable of describing a variety of metadata-usage checking rules that involve XML, annotation, and/or code.}
	\\
	\hline
\end{tabular}

  \vspace{-.5em}
\subsection{Accuracy of \tool}\label{ss:eval-effect-mecheck}

To assess how accurately \tool can detect metadata-related bugs, we created a ground truth dataset of buggy EAs (Section~\ref{sss:data}), defined evaluation metrics (Section~\ref{sss:metrics}), and evaluated \tool using the dataset and metrics (Section~\ref{sss:bug-detect}). 

  \vspace{-.5em}
\subsubsection{A Dataset of Buggy EAs} \label{sss:data}
We mined GitHub~\cite{github} for enterprise applications, using the heuristic-based approach proposed by prior work~\cite{Wen20}. Specifically, we crawled GitHub for any project that contains at least one XML file, whose file path has any of the following keywords: ``Spring", ``security", ``web", and ``WEB-INF". If a project satisfies this requirement, it is likely to be an EA. 
Next, we refined the mined projects by removing irrelevant and redundant projects, getting \totalProject projects.

To create the dataset of buggy EAs, we injected \totalInjection bugs in \totalInjectedProject different projects. To inject bugs for each constraint mentioned in Table~\ref{tab:constraints}, we leveraged the involved metadata items (e.g., ``\codefont{@RunWith(Parameterized.class)}'' or ``\codefont{<property>}'') as keywords to search for three different projects. We manually inspected the relevant metadata usage to ensure that each retrieved project does not violate the corresponding constraint. Then we modified the metadata to intentionally inject a bug for each of the projects. For the 15 rules, we retrieved and modified in total \totalInjectedProject(=15*3) projects. In this way, we injected \totalInjection bugs to \totalInjectedProject distinct open-source projects, making each analyzer inside \tool have 3 bugs to find. We considered these \totalInjection injected bugs as ground truth to evaluate the detection capability of \tool.


   \vspace{-.5em}
\subsubsection{Metrics} \label{sss:metrics}
	We used three metrics to evaluate \tool's effectiveness for bug detection.

\textbf{Precision (P)} measures among all bugs \tool reported, how many of them are true positives:
\begin{equation*}\label{eq:precision}
	    P = \frac{\mbox{\# of correctly reported bugs}}{\mbox{Total \# of bug reports}} 
	\end{equation*}
	
\textbf{Recall (R)} measures among all known bugs, how many of them are reported by \tool:
	\begin{equation*}
	\label{eq:recall}
	    R = \frac{\mbox{\# of correctly reported bugs}}{\mbox{Total \# of known bugs}}
	\end{equation*}

\textbf{F-score (F)} is the harmonic mean of P and R; it reflects the bug-detection accuracy of \tool: 
	\begin{equation*}
	\label{eq:fscore}
	    F = \frac{2 \times P \times R}{P + R} 
	\end{equation*}
        
All metrics mentioned above have values in [0\%, 100\%]; the higher, the better. To compute P and R, we will intersect the set of bugs reported with the ground-truth bug set; any overlap between the two sets captures bugs correctly reported by \tool. Thus, the larger overlap there is, the better.

\subsubsection{Experiment Results} \label{sss:bug-detect}
\tool detected bugs with high precision (\precision), high recall (\recall), and high F-score (\fscore). In total, \tool reported 43 bugs, all of which match the ground truth of injected bugs. However, it missed two bugs related to the rule $r_2$ bean-class-exists, due to the design choice we made when implementing \tool.

  \vspace{-.5em}
\begin{lstlisting}[label={lst:bean-class-exists},caption=The RSL rule of bean-class-exists]
Rule bean-class-exists {
  for (file xml in getXMLs()) {
    if (elementExists(xml, "<bean>")) {
      for (<bean> bean in getElms(xml, "<bean>")) {
        String beanClassFQN = getAttr(bean, "class");
        assert(classExists(beanClassFQN) OR isLibraryClass(beanClassFQN)) {
          msg("Bean class: %s mentioned in bean: %s, does not exist", beanClassFQN, getName(bean)); }}}}}  
\end{lstlisting}

With more details, $r_2$ ensures that if the fully qualified name of a Java class is mentioned as the \codefont{class}-attribute of any \codefont{<bean>}, the class should be either defined by the 
EA-under-analysis or a library on which EA depends. Listing~\ref{lst:bean-class-exists} shows the RSL specification of bean-class-exists, which calls \codefont{callExists(...)} and \codefont{isLibraryClass(...)} on line 6. 
In \tool, \codefont{classExists(...)} is implemented to check whether the pass-in parameter matches any class item extracted from Java code.
The implementation of \codefont{isLibraryClass(...)} defines a list of regular expressions (RegEx) to describe the naming patterns of frequently used library classes, such as ``\codefont{\textasciicircum org\textbackslash.hibernate\textbackslash..+\textdollar}'' for Hibernate classes. We crafted the regular expressions based on (1) our domain knowledge of widely used libraries, and (2) libraries we frequently observed in EAs.
Next, \codefont{isLibraryClass(...)} 
 examines whether a pass-in parameter matches any RegEx to determine if the class is defined by a library. 
We manually injected each of the two missed bugs, by specifying an invalid class name as the \codefont{class}-attribute of \codefont{<bean>}. Ideally, \tool should report both bugs because the \codefont{<bean>}-classes do not exist. However, since the invalid names we specified (e.g., \codefont{org.hibernate.search.hibernate.example.dao.impl.BookDaoImplChanged}) accidentally match predefined RegEx patterns, \tool incorrectly considered them to be valid and failed to locate both bugs.

We could have reduced such false negatives by discarding some RegEx patterns used in function \codefont{isLibraryClass(...)}. However, the downside of doing so is that 
many library classes will fail to match the remaining patterns. Consequently, \tool will wrongly interpret those classes as non-library classes, and report false positives when it also fails to find those classes defined by EA. 
In another word, we may get false positives when trying to overcome such false-negative issues. We prefer generating precise bug reports as developers may not want to get frequently bothered  with false positives. Thus, we designed \tool to include more RegEx patterns and report bugs more precisely, instead of improving recall rates at the cost of sacrificing precision.


\vspace{0.5em}
\noindent\begin{tabular}{|p{13.6cm}|}
	\hline
	\textbf{Finding 2 (Answer to RQ2):} \emph{\tool demonstrated great detection accuracy when being applied to our 45-project dataset with injected bugs. It found bugs with \precision precision, \recall recall, and \fscore F-score. }
	\\
	\hline
\end{tabular}

\subsection{Detection of Metadata-Related Bugs in Real-World Settings}\label{ss:eval-real-world}

To assess how well \tool reveals real bugs, we also created a dataset of \totalRealProject real-world open-source software repositories (Section~\ref{sec:dataset}), and applied \tool to that dataset (Section~\ref{sec:real-setting}). 

\subsubsection{A Dataset of Real-World Software Repositories}\label{sec:dataset}
To create this dataset, we started with the \totalProject projects initially mined from GitHub (see Section~\ref{sss:data}). From those projects, we removed 85 projects that are irrelevant to any of the \totalRule rules implemented in \tool. Namely, if a project does not contain any code/XML/annotation item involved in those rules, we consider it irrelevant. Afterwards, we 
removed the \totalInjection projects used for creating buggy EAs (see Section~\ref{sss:data}), to avoid doing multiple experiments on the same projects.  
{Na\"ively, we could experiment with all program versions{,} for each of the remaining 701 projects, to check whether any version has metadata-related bugs and violates our predefined rules. However, it  would take too much time to analyze all versions of each project,} 
so we decided to only experiment with a subset of the pool.  
To ensure the representatives of our experiment and results, we decided to include projects of distinct scales (i.e., small, medium, and large projects). Thus, we ranked the 701 projects in ascending order of Java file counts, as file counts roughly reflect project sizes. We randomly sampled 10 projects for every 100-project interval, and got \totalRealProject projects. 

In our resulting dataset, each project has 1--987 Java files and 5--1068 total files; the mean value of Java files in each project is 68; the median value of Java files is 24.

\subsubsection{Experiment and Results}\label{sec:real-setting}

We applied \tool to different versions of each selected project, to thoroughly explore whether developers committed any mistakes when maintaining metadata and related Java code. Because many projects have long version histories, it can be very time-consuming to apply \tool to all versions of each project. 
Therefore, to accelerate the bug detection procedure, we developed scripts to filter versions, and focus our experiments on versions that edit any Java or XML file containing code/annotation/XML items relevant to the \totalRule rules. 

Our results show that among the selected \totalRealProject software repositories, \tool reported in total \totalReportedInReal bugs in 21 projects. After manually inspecting the bug reports, we found \totalInteresting reports to be true positives, as they violate either $r_1$ (i.e., xml-path-check) or $r_2$ (i.e., bean-class-exists).   
{In particular, 18 bugs were detected in the first commits of software repositories; 
{99} bugs were found in later commits, meaning that developers accidentally introduced the bugs when they revised software for maintenance.}
{115 of the \totalInteresting bugs were reported together with bugs of the same kind. The existence of multiple bugs in the same program version does not affect our evaluation results, as bugs are analyzed and reported independently. }
{Furthermore, we noticed that \totalRealBugs of the \totalInteresting true positives were already fixed by developers. We found that out by checking later versions of the same projects and by observing revision of those buggy programs.} The remaining 68 bugs had not been fixed yet, so we filed bug reports to contact developers and seek for their feedback. So far, we have not heard from developers for those bug reports.
There are \totalFalsePositives false positives in the \totalReportedInReal bug reports, because the RegEx patterns we used to identify library classes do not cover all the libraries used by EAs.

\begin{table}
\footnotesize
    \centering
        \caption{The \totalRealBugs real bugs later fixed by developers}
    \label{tab:real-bug-fix} \vspace{-1em} 
    \begin{tabular}{r l r R{2cm} r R{3cm}}
\toprule 
    \textbf{Idx}& \textbf{Project Name}&\textbf{\# of Bugs}&\textbf{Ddiff(bug, fix) {(days)}}&\textbf{Vdiff(bug, fix)} &\textbf{Time cost per version ({seconds})}\\ \toprule
    1 & aioweb~\cite{aioweb} & 9 & 1--2 & 1--2&4--6\\ \hline
    2 & angular-js-spring-mybatis~\cite{angular} & 1 & 0 & 1 & 0\\ \hline
    3 & biyam\_repository~\cite{biyam} & 2 &194 & 4&2 \\ \hline
    4 & cv-web~\cite{cv-web} & 2 & 44--47& 1--2&0 \\ \hline
    5 & enterprise-routing-system~\cite{enterprise-routing-system} &2 & 0& 1&5--6 \\ \hline
    6 & FileExplorer~\cite{fe} &1 &0 & 1&3 \\ \hline
    7 & generica~\cite{generica} &2 & 0 &1&2\\ \hline
    8 & I377-esk~\cite{I377} & 1 & 0 &1&7 \\ \hline
    9 & jarvis~\cite{jarvis} & 
    {4}& 9--10 &4--5&2--6\\ \hline
    10 & johnsully83\_groovy~\cite{johnsully} &10 &23&2 &17\\ \hline
    11 & Kognitywistyka~\cite{kognity} & 8 & 0 &1--3 &3--4\\ \hline
    12 & LIBRARY~\cite{library} & 2 & 0 & 1&2\\ \hline
    13 & rop~\cite{rop} & 
    {3} &5--
    {556} & 2--
    {27} &7--9\\ \hline
    14 & ShcUtils~\cite{shcutils} & 1 & 1 & 1&0\\ \hline
    15 & spring-vaadin~\cite{vaadin} & 1 & 10 &1&0\\ \bottomrule
    \end{tabular}
\vspace{-2.em}
\end{table}

Table~\ref{tab:real-bug-fix} presents 
the 
\totalRealBugs bugs that developers later fixed.
All these bugs violate $r_2$---the bean-class-exists rule. Namely, each of the bugs references a nonexistent class when declaring a bean.
In Table~\ref{tab:real-bug-fix}, 
 column \textbf{Idx} shows the index we assigned to each buggy project. \textbf{Project Name} lists the projects where the bugs occurred. \textbf{\# of Bugs} counts the total number of bugs we found in different versions of each project. \textbf{Ddiff(bug, fix)} describes the day difference between committing dates of a bug-fixing version and the related bug-introducing version. Similarly, \textbf{Vdiff(bug, fix)} describes the version difference between the bug-fixing version and buggy version. Let us take the second project angular-js-spring-mybatis~\cite{angular} as an example. In one commit $C_i$ checked in on 
 {Jan 29, 2014}, {developers declared a bean by wrongly referencing a nonexistent class.} In a later commit $C_{i+1}$ checked in on the same day, developers fixed the bug by {correcting the class reference}. Therefore, Vdiff(bug, fix) = (i+1) - i = 1, and Ddiff(bug, fix) = 0. 

For only 
{17} of the \totalRealBugs bugs, developers applied fixes on the same day. However, for the remaining 
{32} bugs, developers applied fixes either on the following day or after quite a long time. 
{15 of the bugs we detected were fixed 1-10 days after the bug-introducing commits; 17 of the bugs were fixed more than 10 days after the bug-introducing commits. 
The largest time difference we observed between a bug-fixing commit and a buggy version is 556 days. 
The phenomena imply the difficulty of revealing those bugs, and the necessity of our approach.}
Additionally, 
{for 24 of the \totalRealBugs bugs, developers applied fixes in the immediately next commit. However, they fixed the remaining bugs after at least two commits. Most interestingly, developers fixed a bug in rop~\cite{rop} after 
{27} commits. 
These observations indicate the great benefits developers can potentially get out of the \tool usage.  
Namely, if developers had used \tool to examine their software before committing program changes, they should have avoided checking in the erroneous program changes, or even have fixed the introduced bugs earlier. 


Finally, \tool has very low runtime cost. We experimented with a laptop that has (1) an Intel i7-8565U CPU with four cores and eight logical processors, and (2) 15.9 GB memory. 
As shown in Table~\ref{tab:real-bug-fix}, when \tool was applied to detect bugs based on the \totalRule rules, it spent no more than 6 seconds on each program version. 

\vspace{0.5em}
\noindent\begin{tabular}{|p{13.6cm}|}
	\hline
	\textbf{Finding 3 (Answer to RQ3):} \emph{\tool demonstrated great effectiveness and high performance when being applied to different versions of 70 real-world open-source EAs. It reported in total \totalReportedInReal bugs; \totalInteresting of the bugs are true positives, \totalRealBugs of which have been already fixed by developers
 }
	\\
	\hline
\end{tabular}
 \vspace{-1em}
\section{Threats to Validity}\label{se:threats}

\paragraph{Threat to External Validity}
We leveraged RSL to express \totalRule rules summarized for Spring and JUnit, and applied \tool to in total 115 (\totalInjectedProject + \totalRealProject) open-source projects for bug detection. The rule set may be limited to our experience with Java frameworks, while the experiment results can be limited to our project datasets. In the future, we would like to include more rules and more projects into our evaluation, or even include close-source projects if possible, so that our findings are more representative. 

\vspace{-.5em}
\paragraph{Threat to Construct Validity} Although we tried our best to manually inspect bug reports, it is possible that our manual analysis are subject to human bias and restricted by our limited domain knowledge. To mitigate the problem, we sent emails to developers who owned the open-source projects and asked whether a reported rule violation makes sense or not. So far, we have not received much feedback from those developers. As we gather more comments from these domain experts, we can further improve the quality of bug detection.

\vspace{-.5em}
\paragraph{Threat to Internal Validity} 
Currently, \tool determines whether a given class $C$ is defined by a library dependency of EA using pattern matching. Namely, if the fully qualified name of $C$ matches a RegEx pattern predefined in \tool, it is considered a library class. However, our RegEx pattern set may not be comprehensive; some library classes may fail the matching process and get wrongly treated as non-library classes. One way to overcome this limitation is to eagerly scan the bytecode of all library dependencies, trying to find an exact match for $C$'s name. However, we decided not to perform such a heavyweight scanning for library classes because based on our experience, many open-source projects do not contain all JAR files for their library dependencies. Such a lack of dependency information will considerably limit the effectiveness of a heavyweight scanning and the applicability of \tool.
In the future, we will investigate more advanced lightweight approaches to mitigate this issue.

{\section{Discussion}}

This section discusses the importance of our \totalRule rules, bug criticality, potential application scope of \tool, the necessity of defining new rules, and a potential alternative design as well as implementation of \tool.

\subsection{Rule Importance and Bug Criticality}

The \totalRule rules we investigated are important, because
 we extracted them from the documentation of Spring and JUnit. All the real bugs we revealed using those rules are critical and severe, as they all trigger runtime instead of compilation errors. It means that without any tool support, developers have to wait till the testing phase, to reveal those rule violations. Additionally, developers spent lots of time revealing some of the bugs we found. As mentioned in Section~\ref{ss:eval-real-world}, 15 of the bugs we detected were fixed 1-10 days after the bug-introducing commits; 17 of the bugs were fixed more than 10 days after the bug-introducing commits. These phenomena imply the difficulty of revealing those bugs, and the necessity of our approach.

According to our experiment for Section~\ref{ss:eval-real-world}, nevertheless, not all rules were violated by real-world projects. One possible reason can be that some developers fixed those violations before checking in commits.
Even if we only observed violations of $r_1$ and $r_2$ in our own real-world experiment, the online resources listed in Table~\ref{tab:rule-and-violate} show that people violated or are likely to violate the remaining rules. Thus, the rules are important as developers tend to violate them.

\vspace{-.5em}
\subsection{The Potential Application Scope of \tool}
So far, we have defined \totalRule RSL rules based on the documentation of Spring and JUnit. However, the application scope of \tool is not limited to these \totalRule rules.
To use \tool, developers can also define their own rules based on other metadata-related constraints~\cite{inspect} or their manual rule extraction from other libraries/frameworks. As XML and annotations have been prevalent configuration methods in various areas and well-known libraries/frameworks (see Table~\ref{tab:area-and-lib}), we believe that \tool can be applied to a wider scope than what is demonstrated in this paper. Essentially, \tool is applicable to arbitrary Java projects that use either XML-based configuration only, annotation-based configuration only, or both XML-based and annotation-based configurations.

\begin{table}
\scriptsize
\begin{minipage}{.53\linewidth}
\caption{The potential areas and libraries/frameworks where \tool is applicable}\label{tab:area-and-lib}
\vspace{-1.5em}
\begin{tabular}{p{2.8cm}l}
\toprule
\textbf{Area} &\textbf{Exemplar Libraries or Frameworks} \\
\toprule
Enterprise Applications & JavaEE \\ \hline
Testing Frameworks &JUnit, TestNG\\ \hline
Dependency Injection and Application Frameworks &Spring Framework, Camel, Guice\\ \hline
Object-Relational Mapping & Hibernate, MyBatis\\ \hline
Web Development & Spring MVC, Struts, JSF\\ \hline
(De)Serialization & Jackson, Gson\\ \hline
Build Tools & Maven, Ant\\ \hline
UI Layouts & Android\\ \hline
Security & Spring Security\\ \hline
Microservice Frameworks & Quarkus, Micronaut \\\bottomrule
\end{tabular}
\end{minipage}
\hfill
\begin{minipage}{.4\linewidth}
\caption{The additional rule violations that get reported or discussed by developers}\label{tab:rule-and-violate}
\vspace{-1.5em}
\centering
\begin{tabular}{l|R{2.5cm}}
\toprule
\textbf{Rule} &\textbf{Violations Reported or Discussed} \\ \toprule
$r_3$     & \cite{r3-violate,r3-violate-2,r3-violate-3} \\ \hline
$r_5$ &\cite{r5-violate}\\ \hline
$r_7$ &\cite{r7-violate}\\ \hline
$r_8$ &\cite{r8-violate, r8-violate-2}\\ \hline
$r_{12}$ &\cite{r12-violate,r12-violate-2}\\ \hline
$r_{13}$ &\cite{r8-violate}\\ \hline
$r_{14}$ &\cite{r14-violate}\\ \hline
$r_{15}$ &\cite{r15-violate}\\ \bottomrule
\end{tabular}
\end{minipage}
\vspace{-1em}
\end{table}

\subsection{The Necessity of Defining New Rules}

We foresee that developers are motivated to define new rules, when they use \tool to examine metadata usage. Three reasons help explain the motivation. 
First, many existing rules~\cite{inspect} are expressible with RSL and checkable by \tool. It indicates a strong need for developers to extend \tool's current 15-rule set to cover those known rules.
Second, prior work~\cite{meng2018secure} shows that when developers asked questions on StackOverflow concerning Spring security usage, the majority of questions are on metadata-based configurations. Such phenomena imply that (1) existing work provides insufficient tool support and (2) some delicate constraints are undocumented. Therefore, it is almost impossible to define a comprehensive ruleset to capture all rules in the wild today. Developers need to extend the ruleset after revealing previously unknown or hidden rules.
Third, as XML and Java annotations are widely used for library/framework configurations, it is likely that future software deriving from these libraries/frameworks will inherit the configuration methods but define new domain-specific rules. Even if we can define a comprehensive ruleset for \tool today, as the time goes, new domain-specific rules appear; developers still need to expand \tool's ruleset  to cover those rules. Therefore, it is necessary and important to define RSL for rule definition.

\subsection{A Potential Alternative Design and Implementation of \tool}
Existing static analysis tools like PMD~\cite{pmd} are created to detect bugs in Java code. Some readers may wonder why we did not extend those tools to define metadata-related rules. We thought about that option but decided to define our own DSL and create a new tool, mainly because the metadata-related rules we focus on are so unique that existing tools barely handle them. Take PMD as an example. PMD does not examine any of the rules listed in this paper. To extend the ruleset of PMD for XML analysis, people have to learn and use XPath---another DSL---to define queries on XML documents. However, XPath does not support queries on source code or annotations. It has a narrower scope than RSL. Furthermore, PMD is complex, with lots of implementation irrelevant to our focus, which can make our tool development on top of PMD very time-consuming and error-prone. To (1) avoid dealing with the complexity of PMD and (2) quickly prototype our research, we created \tool without reusing PMD or XPath.
\section{Related Work}\label{se:related}
The related work of our research includes DSLs defined for metadata usage checking, detection of metadata-related bugs, and configuration debugging.

  \vspace{-.5em}
\subsection{Domain-Specific Languages (DSLs) Defined for Metadata-Usage Checking}

People invented DSLs to check and/or fix the usage of metadata (i.e., XML and annotations)~\cite{
benzaken2003cduce,chamberlin2002xquery,hosoya2003xduce,darwin2009annabot,eichberg2005using,Song12,noguera2008annotation}. For instance, XQuery is a widely used query and functional programming language that queries and transforms collections of structured or unstructured data in XML documents~\cite{chamberlin2002xquery}. Similarly, CDuce~\cite{benzaken2003cduce} and XDuce~\cite{hosoya2003xduce} are independently developed DSLs for XML processing. 
To validate Java annotation usage, 
Eichberg et al.~created a DSL for users to define constraints~\cite{eichberg2005using}. To check user-specified constraints, the researchers automatically converted Java bytecode to XML documents, and converted constraints to XQuery path expressions.  Darwin~\cite{darwin2009annabot} and Noguera et al.~\cite{noguera2008annotation} also  defined distinct DSLs for users to specify and  validate the constraints on annotation usage. 
All DSLs mentioned above only focus on one type of metadata (i.e., either XML or annotations), but not on both types or on the relations between them.


Song and Tilevich created (1) MIL to express metadata invariants, and (2) the language implementation/engine to 
examine code-annotation relations (i.e., relations between Java code and annotations) and code-XML relations (i.e., relations between code and XML)~\cite{Song12}. We tried to run MIL engine in our experiments, but without success. RSL is different from MIL in three ways. First, it is an imperative instead of declarative language, so users have more flexibility in controlling how items are extracted, enumerated, filtered, and checked. 
Second, RSL has stronger expressiveness. It can express rules to examine 
XML-annotation-code constraints and XML-annotation constraints, but we did not find MIL capable of doing that. Third, RSL can  have better performance, because we applied specialized optimizations in \tool to eliminate repetitive computation but MIL engine was unoptimized.

  \vspace{-.5em}
\subsection{Detection of Metadata-Related Bugs}

Researchers created tools to automatically detect metadata-related bugs~\cite{Noguera2012,Wen20,Nuryyev2022,Zhang2024}.
For instance, Wen et al.~created XEDITOR to automatically infer and apply def-use like configuration couplings in XML files~\cite{Wen20}. Specifically, XEDITOR extracts XML entity pairs that (i) frequently coexist in the same files and (ii) hold the same data at least once; it then applies customized association rule mining to infer def-use like couplings ``A$\rightarrow$B'' between entities. For bug detection, given a new XML file, XEDITOR checks whether the file violates any couplings; if so, XEDITOR reports the violation(s). 
Noguera et al.~created an extension of the Eclipse IDE's refactoring engine, to check whether any Java code refactoring violates the
correspondence constraints between existing code and annotations~\cite{Noguera2012}. 
Nuryyev et al.~devised a frequent-itemset based pattern mining approach to mine annotation-usage rules for MicroProfile, an open-source Java microservice framework~\cite{Nuryyev2022}. 
By scanning 533 MicroProfile projects for violations of 12 of the mined rules, the researchers found 100 violations of 5 mined rules in 16 projects. 
Zhang et al.~noticed that existing static program analyzers are unaware of the semantic changes introduced by annotations, and consequently can produce imprecise analysis results~\cite{Zhang2024}. Thus, they conducted a study of annotation-induced faults (AIF) by analyzing 246 issues in 6 open-source and popular static analysis (i.e., PMD, SpotBugs, CheckStyle, Infer, SonarQube, and Soot). Based on their findings in the study, the researchers created a tool to generate new tests for static analyzers, and revealed 43 new faults, 20 of which have been fixed.

Our research complements prior work in two ways. First, it examines correspondence constraints in a wider scope, by querying and checking on the content correspondence in XML files, Java code, and annotations; nevertheless, existing tools only examine correspondence in one type of metadata. 
No existing tool examines the content constraints between XML and annotations,  but \tool does so.
Second, RSL enables people (e.g., library developers) to freely express domain-specific constraints for XML-based and annotation-based configurations. This feature is especially important when some domain-specific constraints are not detectable for any mining tool.
Furthermore, when a constraint is mined by existing tools, people can easily extend \tool by defining an RSL specification for it, to embed the mined knowledge into automatic metadata analysis.



  \vspace{-.5em}
\subsection{Configuration Debugging}
Some tools were built to diagnose or fix software configuration errors~\cite{Attariyan:2008:UCD,Attariyan:2010:ACT,Rabkin:2011:PPC,Zhang:2013:ADS,Weiss:2017:TIS,Oh2021}. 
For instance, Attariyan et al.~\cite{Attariyan:2008:UCD} and Zhang et al.~\cite{Zhang:2013:ADS} separately created tools to record predicates that may be affected by configuration options, collect the execution profiles of a program's correct and undesired runs, and compare the behavioral differences between the two types of runs to diagnose configuration errors. 
Rabkin and Katz created a static analysis-based tool to help users debug configuration errors in software~\cite{Rabkin:2011:PPC}. 
The tool tracks the flow of configuration labels through a program: labels get introduced via configuration reads, and propagated via assignment and library calls. 
Weiss et al.~built an approach to generalize system configuration repairs for certain types of machines, from the shell commands developers entered to update one machine~\cite{Weiss:2017:TIS}. 
Oh et al.~introduce a model checking framework for building Kconfig static analysis tools~\cite{Oh2021}. The configuration specification language, Kconfig, is defined to prevent invalid configurations of the Linux kernel from being built. 
To detect bugs in Kconfig specifications, Oh et al.~created a symbolic evaluator \codefont{kclause} to model Kconfig specifications, and a tool to find bugs in \codefont{kclause} models. 
The configuration files examined by these tools are irrelevant to XML documents or Java annotations, so they cannot detect the bugs we  focus on. 

\vspace{-.5em}
\section{Conclusion}\label{se:conclusion}

Similar to 
proper code implementation, correct metadata usage is 
essential to software quality. 
Despite
the importance of high-quality metadata usage, widely used program analysis techniques (e.g., WALA~\cite{wala} and Soot~\cite{soot}) 
provide little support for
metadata debugging.
Some existing tools can help developers identify certain metadata bugs, but the tool support is quite limited. 
In this paper, we developed RSL---a domain-specific language for describing metadata-usage checking rules, and built \tool---a tool to analyze programs based on RSL rules. Compared with existing tools, our approach is unique in three aspects. First, RSL enables library/framework developers to freely specify domain-specific constraints on the relations among code, annotations, and XML items; thus, it empowers \tool to enforce diverse rules in a wider scope and equips \tool with a strong power extensibility. 
Second, \tool extracts program data from diverse software artifacts (e.g., Java source files and XML files), and reasons about the relationship between the extracted data to detect bugs. It demonstrates a great way of integrating source code analysis with metadata analysis. Third, \tool applies optimizations to reduce redundant data processing, when trying to establish pattern matching between a given program and multiple bug patterns described by RSL rules. No prior work pays such a close attention to the time cost of metadata-related bug detection.

There is still significant space for future improvements in static analysis for metadata usage. 
In the future, we will investigate more domain-specific rules posed by different library frameworks, and extend \tool to conduct more advanced synergetic analysis among code and metadata.



\vspace{-.5em}

\section*{Data Availability}
The data and programs are available at \url{https://doi.org/10.5281/zenodo.15205192}.

\section*{Acknowledgments}
We thank all reviewers for their valuable feedback. This work was partially funded by NSF-2006278, NSF-1929701, NSF-1845446, NSF-2007718, and CCI. 

\bibliographystyle{ACM-Reference-Format}
\bibliography{dsl-icse25}


\begin{thebibliography}{72}


\ifx \showCODEN    \undefined \def \showCODEN     #1{\unskip}     \fi
\ifx \showDOI      \undefined \def \showDOI       #1{#1}\fi
\ifx \showISBNx    \undefined \def \showISBNx     #1{\unskip}     \fi
\ifx \showISBNxiii \undefined \def \showISBNxiii  #1{\unskip}     \fi
\ifx \showISSN     \undefined \def \showISSN      #1{\unskip}     \fi
\ifx \showLCCN     \undefined \def \showLCCN      #1{\unskip}     \fi
\ifx \shownote     \undefined \def \shownote      #1{#1}          \fi
\ifx \showarticletitle \undefined \def \showarticletitle #1{#1}   \fi
\ifx \showURL      \undefined \def \showURL       {\relax}        \fi
\providecommand\bibfield[2]{#2}
\providecommand\bibinfo[2]{#2}
\providecommand\natexlab[1]{#1}
\providecommand\showeprint[2][]{arXiv:#2}

\bibitem[con(2012)]%
        {confuse_error}
 \bibinfo{year}{2012}\natexlab{}.
\newblock \bibinfo{title}{Securing REST urls with Spring}.
\newblock
\newblock
\urldef\tempurl%
\url{https://stackoverflow.com/questions/13836451/securing-rest-urls-with-spring}
\showURL{%
\tempurl}


\bibitem[r3-(2014)]%
        {r3-violate-3}
 \bibinfo{year}{2014}\natexlab{}.
\newblock \bibinfo{title}{{BeanCreationException in spring controller}}.
\newblock \bibinfo{howpublished}{\url{https://stackoverflow.com/questions/21520827/beancreationexception-in-spring-controller?rq=3}}.
\newblock


\bibitem[r5-(2014)]%
        {r5-violate}
 \bibinfo{year}{2014}\natexlab{}.
\newblock \bibinfo{title}{{"Could not resolve matching constructor" error when passing in constructor-arg for child class}}.
\newblock \bibinfo{howpublished}{\url{https://stackoverflow.com/questions/21583399/could-not-resolve-matching-constructor-error-when-passing-in-constructor-arg-f}}.
\newblock


\bibitem[ora(2014)]%
        {oracle2014xmlparsers}
 \bibinfo{year}{2014}\natexlab{}.
\newblock \bibinfo{title}{Package javax.xml.parsers for processing of XML documents}.
\newblock \bibinfo{howpublished}{Oracle Java API Documentation}.
\newblock
\urldef\tempurl%
\url{https://docs.oracle.com/javase/8/docs/api/index.html?javax/xml/parsers/package-summary.html}
\showURL{%
\tempurl}


\bibitem[r3-(2016)]%
        {r3-violate-2}
 \bibinfo{year}{2016}\natexlab{}.
\newblock \bibinfo{title}{{Could not resolve matching constructor.Ambiguity issue with Spring dependency injection}}.
\newblock \bibinfo{howpublished}{\url{https://stackoverflow.com/questions/37648984/could-not-resolve-matching-constructor-ambiguity-issue-with-spring-dependency-in?rq=3}}.
\newblock


\bibitem[mis(2016)]%
        {misconfiguration}
 \bibinfo{year}{2016}\natexlab{}.
\newblock \bibinfo{title}{{Custom Authentication Filters in multiple HttpSecurity objects using Java Config}}.
\newblock \bibinfo{howpublished}{\url{https://stackoverflow.com/questions/37304211/custom-authentication-filters-in-multiple-httpsecurity-objects-using-java-config}}.
\newblock


\bibitem[r15(2016)]%
        {r15-violate}
 \bibinfo{year}{2016}\natexlab{}.
\newblock \bibinfo{title}{{org.springframework.beans.factory.BeanCreationException in Spring boot application}}.
\newblock \bibinfo{howpublished}{\url{https://stackoverflow.com/questions/37938942/org-springframework-beans-factory-beancreationexception-in-spring-boot-applicati}}.
\newblock


\bibitem[spr(2016)]%
        {springSecurity}
 \bibinfo{year}{2016}\natexlab{}.
\newblock \bibinfo{title}{Spring security JDK based proxy issue while using @Secured annotation on Controller method}.
\newblock
\newblock
\urldef\tempurl%
\url{https://stackoverflow.com/questions/35860442/spring-security-jdk-based-proxy-issue-while-using-secured-annotation-on-control}
\showURL{%
\tempurl}


\bibitem[sec(2016)]%
        {security-vul}
 \bibinfo{year}{2016}\natexlab{}.
\newblock \bibinfo{title}{{Spring security JDK based proxy issue while using @Secured annotation on Controller method}}.
\newblock \bibinfo{howpublished}{\url{https://stackoverflow.com/questions/35860442/spring-security-jdk-based-proxy-issue-while-using-secured-annotation-on-control}}.
\newblock


\bibitem[r7-(2016)]%
        {r7-violate}
 \bibinfo{year}{2016}\natexlab{}.
\newblock \bibinfo{title}{{XML Based Configuration using Spring and Hibernate}}.
\newblock \bibinfo{howpublished}{\url{https://stackoverflow.com/questions/39891823/xml-based-configuration-using-spring-and-hibernate}}.
\newblock


\bibitem[r12(2019)]%
        {r12-violate}
 \bibinfo{year}{2019}\natexlab{}.
\newblock \bibinfo{title}{{JUnit No Runnable Methods}}.
\newblock \bibinfo{howpublished}{\url{https://examples.javacodegeeks.com/java-development/core-java/junit/junit-no-runnable-methods/}}.
\newblock


\bibitem[r8-(2021)]%
        {r8-violate-2}
 \bibinfo{year}{2021}\natexlab{}.
\newblock \bibinfo{title}{{Getting error when running the code}}.
\newblock \bibinfo{howpublished}{\url{https://www.qtpselenium.com/selenium-training/forum/7159/getting-error-when-running-the-code}}.
\newblock


\bibitem[jav(2021a)]%
        {javaEE}
 \bibinfo{year}{2021}\natexlab{a}.
\newblock \bibinfo{title}{{Java EE at a Glance}}.
\newblock \bibinfo{howpublished}{\url{https://www.oracle.com/java/technologies/java-ee-glance.html}}.
\newblock


\bibitem[spr(2021a)]%
        {spring}
 \bibinfo{year}{2021}\natexlab{a}.
\newblock \bibinfo{title}{Spring}.
\newblock \bibinfo{howpublished}{\url{https://spring.io}}.
\newblock


\bibitem[spr(2021b)]%
        {spring-anno}
 \bibinfo{year}{2021}\natexlab{b}.
\newblock \bibinfo{title}{{Spring - Annotation Based Configuration}}.
\newblock \bibinfo{howpublished}{\url{https://www.tutorialspoint.com/spring/spring_annotation_based_configuration.htm}}.
\newblock


\bibitem[spr(2021c)]%
        {spring-doc}
 \bibinfo{year}{2021}\natexlab{c}.
\newblock \bibinfo{title}{{Spring Framework Documentation}}.
\newblock \bibinfo{howpublished}{\url{https://docs.spring.io/spring-framework/docs/current/reference/html/}}.
\newblock


\bibitem[spr(2021d)]%
        {spring-tutorial}
 \bibinfo{year}{2021}\natexlab{d}.
\newblock \bibinfo{title}{{Spring Tutorial}}.
\newblock \bibinfo{howpublished}{\url{https://www.tutorialspoint.com/spring/index.htm}}.
\newblock


\bibitem[jav(2021b)]%
        {javaee-anno}
 \bibinfo{year}{2021}\natexlab{b}.
\newblock \bibinfo{title}{{Using Java EE Annotations and Dependency Injection}}.
\newblock \bibinfo{howpublished}{\url{https://docs.oracle.com/cd/E11035_01/wls100/programming/annotate_dependency.html}}.
\newblock


\bibitem[DD_(2021)]%
        {DD_javaEE}
 \bibinfo{year}{2021}\natexlab{}.
\newblock \bibinfo{title}{XML Deployment Descriptors}.
\newblock \bibinfo{howpublished}{\url{https://docs.oracle.com/cd/A91202_01/901_doc/java.901/a90188/xml.htm}}.
\newblock


\bibitem[xml(2021)]%
        {xml-elements}
 \bibinfo{year}{2021}\natexlab{}.
\newblock \bibinfo{title}{{XML Elements}}.
\newblock \bibinfo{howpublished}{\url{https://www.geeksforgeeks.org/xml-elements/}}.
\newblock


\bibitem[ini(2022)]%
        {init-destroy}
 \bibinfo{year}{2022}\natexlab{}.
\newblock \bibinfo{title}{{Spring -- init() and destroy() Methods with Example}}.
\newblock \bibinfo{howpublished}{\url{https://www.geeksforgeeks.org/spring-init-and-destroy-methods-with-example/}}.
\newblock


\bibitem[r3-(2023)]%
        {r3-violate}
 \bibinfo{year}{2023}\natexlab{}.
\newblock \bibinfo{title}{{Ambiguous argument values for parameter of type [int]}}.
\newblock \bibinfo{howpublished}{\url{https://blog.csdn.net/qq_49676677/article/details/119714497}}.
\newblock


\bibitem[I37(2024)]%
        {I377}
 \bibinfo{year}{2024}\natexlab{}.
\newblock \bibinfo{title}{{}}.
\newblock \bibinfo{howpublished}{\url{https://github.com/hannesnolvak/I377-esk}}.
\newblock


\bibitem[ang(2024)]%
        {angular}
 \bibinfo{year}{2024}\natexlab{}.
\newblock \bibinfo{title}{{angular-js-spring-mybatis}}.
\newblock \bibinfo{howpublished}{\url{https://github.com/rurutia/angular-js-spring-mybatis}}.
\newblock


\bibitem[fe(2024)]%
        {fe}
 \bibinfo{year}{2024}\natexlab{}.
\newblock \bibinfo{title}{{brianleesg1/FileExplorer}}.
\newblock \bibinfo{howpublished}{{https://github.com/brianleesg1/FileExplorer}}.
\newblock


\bibitem[lib(2024)]%
        {library}
 \bibinfo{year}{2024}\natexlab{}.
\newblock \bibinfo{title}{{dawe73/LIBRARY}}.
\newblock \bibinfo{howpublished}{\url{https://github.com/dawe73/LIBRARY}}.
\newblock


\bibitem[biy(2024)]%
        {biyam}
 \bibinfo{year}{2024}\natexlab{}.
\newblock \bibinfo{title}{{gedkang / biyam\_repository}}.
\newblock \bibinfo{howpublished}{\url{https://github.com/gedkang/biyam_repository}}.
\newblock


\bibitem[git(2024)]%
        {github}
 \bibinfo{year}{2024}\natexlab{}.
\newblock \bibinfo{title}{Github}.
\newblock
\newblock
\urldef\tempurl%
\url{https://github.com/}
\showURL{%
\tempurl}


\bibitem[aio(2024)]%
        {aioweb}
 \bibinfo{year}{2024}\natexlab{}.
\newblock \bibinfo{title}{{hopestar720/aioweb}}.
\newblock \bibinfo{howpublished}{\url{https://github.com/hopestar720/aioweb}}.
\newblock


\bibitem[how(2024)]%
        {how-to-junit}
 \bibinfo{year}{2024}\natexlab{}.
\newblock \bibinfo{title}{{How to create JUnit Test Suite? (with Examples)}}.
\newblock \bibinfo{howpublished}{\url{https://www.browserstack.com/guide/junit-test-suite}}.
\newblock


\bibitem[jav(2024)]%
        {javaCC_page}
 \bibinfo{year}{2024}\natexlab{}.
\newblock \bibinfo{title}{JavaCC, The most popular parser generator for use with Java applications.}
\newblock
\newblock
\urldef\tempurl%
\url{https://javacc.github.io/javacc/}
\showURL{%
\tempurl}


\bibitem[joh(2024)]%
        {johnsully}
 \bibinfo{year}{2024}\natexlab{}.
\newblock \bibinfo{title}{{johnsully83/johnsully83\_groovy}}.
\newblock \bibinfo{howpublished}{{https://github.com/johnsully83/johnsully83\_groovy}}.
\newblock


\bibitem[jun(2024a)]%
        {junit-tutorial}
 \bibinfo{year}{2024}\natexlab{a}.
\newblock \bibinfo{title}{{JUnit - Suite Test}}.
\newblock \bibinfo{howpublished}{\url{https://www.tutorialspoint.com/junit/junit_suite_test.htm}}.
\newblock


\bibitem[rop(2024)]%
        {rop}
 \bibinfo{year}{2024}\natexlab{}.
\newblock \bibinfo{title}{{liuzhaomincoding / rop}}.
\newblock \bibinfo{howpublished}{\url{https://github.com/liuzhaomincoding/rop}}.
\newblock


\bibitem[kog(2024)]%
        {kognity}
 \bibinfo{year}{2024}\natexlab{}.
\newblock \bibinfo{title}{{m2gikbb/Kognitywistyka}}.
\newblock \bibinfo{howpublished}{\url{https://github.com/m2gikbb/Kognitywistyka}}.
\newblock


\bibitem[vaa(2024)]%
        {vaadin}
 \bibinfo{year}{2024}\natexlab{}.
\newblock \bibinfo{title}{{mcgray/spring-vaadin}}.
\newblock \bibinfo{howpublished}{\url{https://github.com/mcgray/spring-vaadin}}.
\newblock


\bibitem[ins(2024)]%
        {inspect}
 \bibinfo{year}{2024}\natexlab{}.
\newblock \bibinfo{title}{{New Inspections in This Release | Inspectopedia Documentation}}.
\newblock \bibinfo{howpublished}{\url{https://www.jetbrains.com/help/inspectopedia/}}.
\newblock


\bibitem[jun(2024b)]%
        {junit-doc}
 \bibinfo{year}{2024}\natexlab{b}.
\newblock \bibinfo{title}{{Parameterized (JUnit API)}}.
\newblock \bibinfo{howpublished}{\url{https://junit.org/junit4/javadoc/4.12/org/junit/runners/Parameterized.html}}.
\newblock


\bibitem[r8-(2024)]%
        {r8-violate}
 \bibinfo{year}{2024}\natexlab{}.
\newblock \bibinfo{title}{{Parameterized test class without data provider method}}.
\newblock \bibinfo{howpublished}{\url{https://www.jetbrains.com.cn/en-us/help/inspectopedia/ParameterizedParametersStaticCollection.html}}.
\newblock


\bibitem[jar(2024)]%
        {jarvis}
 \bibinfo{year}{2024}\natexlab{}.
\newblock \bibinfo{title}{{ReeverPD/jarvis}}.
\newblock \bibinfo{howpublished}{\url{https://github.com/ReeverPD/jarvis}}.
\newblock


\bibitem[shc(2024)]%
        {shcutils}
 \bibinfo{year}{2024}\natexlab{}.
\newblock \bibinfo{title}{{ShcUtils}}.
\newblock \bibinfo{howpublished}{\url{https://github.com/535521469/ShcUtils}}.
\newblock


\bibitem[cv-(2024)]%
        {cv-web}
 \bibinfo{year}{2024}\natexlab{}.
\newblock \bibinfo{title}{{smyrouf/cv-web}}.
\newblock \bibinfo{howpublished}{\url{https://github.com/smyrouf/cv-web}}.
\newblock


\bibitem[sof(2024)]%
        {software-testing}
 \bibinfo{year}{2024}\natexlab{}.
\newblock \bibinfo{title}{{SOFTWARE TESTING: GETTING STARTED WITH ECLIPSE AND JUNIT (JUNIT 3)}}.
\newblock \bibinfo{howpublished}{\url{https://www.inf.ed.ac.uk/teaching/courses/st/2010-2011/tutorials/tutorial1j3.html}}.
\newblock


\bibitem[soo(2024)]%
        {soot}
 \bibinfo{year}{2024}\natexlab{}.
\newblock \bibinfo{title}{{Soot}}.
\newblock \bibinfo{howpublished}{\url{https://github.com/soot-oss/soot}}.
\newblock


\bibitem[ent(2024)]%
        {enterprise-routing-system}
 \bibinfo{year}{2024}\natexlab{}.
\newblock \bibinfo{title}{{sovcn/enterprise-routing-system}}.
\newblock \bibinfo{howpublished}{\url{https://github.com/sovcn/enterprise-routing-system}}.
\newblock


\bibitem[gen(2024)]%
        {generica}
 \bibinfo{year}{2024}\natexlab{}.
\newblock \bibinfo{title}{{ui-kreinhard/generica}}.
\newblock \bibinfo{howpublished}{\url{https://github.com/ui-kreinhard/generica}}.
\newblock


\bibitem[r14(2024)]%
        {r14-violate}
 \bibinfo{year}{2024}\natexlab{}.
\newblock \bibinfo{title}{{Unresolved file references in @ImportResource locations}}.
\newblock \bibinfo{howpublished}{\url{https://www.jetbrains.com/help/inspectopedia/SpringImportResource.html}}.
\newblock


\bibitem[wal(2024)]%
        {wala}
 \bibinfo{year}{2024}\natexlab{}.
\newblock \bibinfo{title}{WALA}.
\newblock \bibinfo{howpublished}{\url{https://github.com/wala/WALA}}.
\newblock


\bibitem[r12(2025)]%
        {r12-violate-2}
 \bibinfo{year}{2025}\natexlab{}.
\newblock \bibinfo{title}{{java lang exception no runnable methods}}.
\newblock \bibinfo{howpublished}{\url{https://www.javatpoint.com/java-lang-exception-no-runnable-methods}}.
\newblock


\bibitem[pmd(2025)]%
        {pmd}
 \bibinfo{year}{2025}\natexlab{}.
\newblock \bibinfo{title}{{PMD}}.
\newblock \bibinfo{howpublished}{\url{https://pmd.github.io}}.
\newblock


\bibitem[Attariyan and Flinn(2008)]%
        {Attariyan:2008:UCD}
\bibfield{author}{\bibinfo{person}{Mona Attariyan} {and} \bibinfo{person}{Jason Flinn}.} \bibinfo{year}{2008}\natexlab{}.
\newblock \showarticletitle{Using Causality to Diagnose Configuration Bugs}. In \bibinfo{booktitle}{\emph{USENIX 2008 Annual Technical Conference}} (Boston, Massachusetts) \emph{(\bibinfo{series}{ATC'08})}. \bibinfo{publisher}{USENIX Association}, \bibinfo{address}{Berkeley, CA, USA}, \bibinfo{pages}{281--286}.
\newblock
\urldef\tempurl%
\url{http://dl.acm.org/citation.cfm?id=1404014.1404037}
\showURL{%
\tempurl}


\bibitem[Attariyan and Flinn(2010)]%
        {Attariyan:2010:ACT}
\bibfield{author}{\bibinfo{person}{Mona Attariyan} {and} \bibinfo{person}{Jason Flinn}.} \bibinfo{year}{2010}\natexlab{}.
\newblock \showarticletitle{Automating Configuration Troubleshooting with Dynamic Information Flow Analysis}. In \bibinfo{booktitle}{\emph{Proceedings of the 9th USENIX Conference on Operating Systems Design and Implementation}} (Vancouver, BC, Canada) \emph{(\bibinfo{series}{OSDI'10})}. \bibinfo{publisher}{USENIX Association}, \bibinfo{address}{Berkeley, CA, USA}, \bibinfo{pages}{237--250}.
\newblock
\urldef\tempurl%
\url{http://dl.acm.org/citation.cfm?id=1924943.1924960}
\showURL{%
\tempurl}


\bibitem[Benzaken et~al\mbox{.}(2003)]%
        {benzaken2003cduce}
\bibfield{author}{\bibinfo{person}{V{\'e}ronique Benzaken}, \bibinfo{person}{Giuseppe Castagna}, {and} \bibinfo{person}{Alain Frisch}.} \bibinfo{year}{2003}\natexlab{}.
\newblock \showarticletitle{CDuce: an XML-centric general-purpose language}.
\newblock \bibinfo{journal}{\emph{ACM SIGPLAN Notices}} \bibinfo{volume}{38}, \bibinfo{number}{9} (\bibinfo{year}{2003}), \bibinfo{pages}{51--63}.
\newblock


\bibitem[Chamberlin(2002)]%
        {chamberlin2002xquery}
\bibfield{author}{\bibinfo{person}{Don Chamberlin}.} \bibinfo{year}{2002}\natexlab{}.
\newblock \showarticletitle{XQuery: An XML query language}.
\newblock \bibinfo{journal}{\emph{IBM systems journal}} \bibinfo{volume}{41}, \bibinfo{number}{4} (\bibinfo{year}{2002}), \bibinfo{pages}{597--615}.
\newblock


\bibitem[Darwin(2009)]%
        {darwin2009annabot}
\bibfield{author}{\bibinfo{person}{Ian Darwin}.} \bibinfo{year}{2009}\natexlab{}.
\newblock \showarticletitle{Annabot: A static verifier for java annotation usage}.
\newblock \bibinfo{journal}{\emph{Advances in Software Engineering}}  \bibinfo{volume}{2010} (\bibinfo{year}{2009}).
\newblock


\bibitem[Eichberg et~al\mbox{.}(2005)]%
        {eichberg2005using}
\bibfield{author}{\bibinfo{person}{Michael Eichberg}, \bibinfo{person}{Thorsten Sch{\"a}fer}, {and} \bibinfo{person}{Mira Mezini}.} \bibinfo{year}{2005}\natexlab{}.
\newblock \showarticletitle{Using annotations to check structural properties of classes}. In \bibinfo{booktitle}{\emph{International Conference on Fundamental Approaches to Software Engineering}}. Springer, \bibinfo{pages}{237--252}.
\newblock


\bibitem[Hosoya and Pierce(2003)]%
        {hosoya2003xduce}
\bibfield{author}{\bibinfo{person}{Haruo Hosoya} {and} \bibinfo{person}{Benjamin~C Pierce}.} \bibinfo{year}{2003}\natexlab{}.
\newblock \showarticletitle{XDuce: A statically typed XML processing language}.
\newblock \bibinfo{journal}{\emph{ACM Transactions on Internet Technology (TOIT)}} \bibinfo{volume}{3}, \bibinfo{number}{2} (\bibinfo{year}{2003}), \bibinfo{pages}{117--148}.
\newblock


\bibitem[H\"ursch and Lopes(1995)]%
        {Hursch95separationof}
\bibfield{author}{\bibinfo{person}{Walter~L. H\"ursch} {and} \bibinfo{person}{Cristina~Videira Lopes}.} \bibinfo{year}{1995}\natexlab{}.
\newblock \bibinfo{booktitle}{\emph{Separation of Concerns}}.
\newblock \bibinfo{type}{{T}echnical {R}eport}.
\newblock


\bibitem[{JavaParser}({[n.\,d.]})]%
        {javaparser}
\bibfield{author}{\bibinfo{person}{{JavaParser}}.} \bibinfo{year}{[n.\,d.]}\natexlab{}.
\newblock \bibinfo{title}{JavaParser - Tools for your Java code}.
\newblock \bibinfo{howpublished}{\url{https://javaparser.org/}}.
\newblock
\newblock
\shownote{Accessed: 2024-07-21}.


\bibitem[Meng et~al\mbox{.}(2018)]%
        {meng2018secure}
\bibfield{author}{\bibinfo{person}{Na Meng}, \bibinfo{person}{Stefan Nagy}, \bibinfo{person}{Danfeng Yao}, \bibinfo{person}{Wenjie Zhuang}, {and} \bibinfo{person}{Gustavo~Arango Argoty}.} \bibinfo{year}{2018}\natexlab{}.
\newblock \showarticletitle{Secure coding practices in java: Challenges and vulnerabilities}. In \bibinfo{booktitle}{\emph{Proceedings of the 40th International Conference on Software Engineering}}. \bibinfo{pages}{372--383}.
\newblock


\bibitem[Noguera and Duchien(2008)]%
        {noguera2008annotation}
\bibfield{author}{\bibinfo{person}{Carlos Noguera} {and} \bibinfo{person}{Laurence Duchien}.} \bibinfo{year}{2008}\natexlab{}.
\newblock \showarticletitle{Annotation framework validation using domain models}. In \bibinfo{booktitle}{\emph{European Conference on Model Driven Architecture-Foundations and Applications}}. Springer, \bibinfo{pages}{48--62}.
\newblock


\bibitem[Noguera et~al\mbox{.}(2012)]%
        {Noguera2012}
\bibfield{author}{\bibinfo{person}{Carlos Noguera}, \bibinfo{person}{Andy Kellens}, \bibinfo{person}{Coen De~Roover}, {and} \bibinfo{person}{Viviane Jonckers}.} \bibinfo{year}{2012}\natexlab{}.
\newblock \showarticletitle{Refactoring in the presence of annotations}. In \bibinfo{booktitle}{\emph{2012 28th IEEE International Conference on Software Maintenance (ICSM)}}. \bibinfo{pages}{337--346}.
\newblock
\urldef\tempurl%
\url{https://doi.org/10.1109/ICSM.2012.6405291}
\showDOI{\tempurl}


\bibitem[Nuryyev et~al\mbox{.}(2022)]%
        {Nuryyev2022}
\bibfield{author}{\bibinfo{person}{Batyr Nuryyev}, \bibinfo{person}{Ajay Kumar~Jha}, \bibinfo{person}{Sarah Nadi}, \bibinfo{person}{Yee-Kang Chang}, \bibinfo{person}{Emily Jiang}, {and} \bibinfo{person}{Vijay Sundaresan}.} \bibinfo{year}{2022}\natexlab{}.
\newblock \showarticletitle{Mining Annotation Usage Rules: A Case Study with MicroProfile}. In \bibinfo{booktitle}{\emph{2022 IEEE International Conference on Software Maintenance and Evolution (ICSME)}}. \bibinfo{pages}{553--562}.
\newblock
\urldef\tempurl%
\url{https://doi.org/10.1109/ICSME55016.2022.00075}
\showDOI{\tempurl}


\bibitem[Oh et~al\mbox{.}(2021)]%
        {Oh2021}
\bibfield{author}{\bibinfo{person}{Jeho Oh}, \bibinfo{person}{Necip~Faz\i{}l Y\i{}ld\i{}ran}, \bibinfo{person}{Julian Braha}, {and} \bibinfo{person}{Paul Gazzillo}.} \bibinfo{year}{2021}\natexlab{}.
\newblock \showarticletitle{Finding broken Linux configuration specifications by statically analyzing the Kconfig language}. In \bibinfo{booktitle}{\emph{Proceedings of the 29th ACM Joint Meeting on European Software Engineering Conference and Symposium on the Foundations of Software Engineering}} (Athens, Greece) \emph{(\bibinfo{series}{ESEC/FSE 2021})}. \bibinfo{publisher}{Association for Computing Machinery}, \bibinfo{address}{New York, NY, USA}, \bibinfo{pages}{893–905}.
\newblock
\showISBNx{9781450385626}
\urldef\tempurl%
\url{https://doi.org/10.1145/3468264.3468578}
\showDOI{\tempurl}


\bibitem[Oracle(2021)]%
        {ea-development}
\bibfield{author}{\bibinfo{person}{Oracle}.} \bibinfo{year}{2021}\natexlab{}.
\newblock \bibinfo{title}{{Overview of Enterprise Applications}}.
\newblock \bibinfo{howpublished}{\url{https://docs.oracle.com/javaee/6/firstcup/doc/gcrky.html}}.
\newblock


\bibitem[Rabkin and Katz(2011)]%
        {Rabkin:2011:PPC}
\bibfield{author}{\bibinfo{person}{Ariel Rabkin} {and} \bibinfo{person}{Randy Katz}.} \bibinfo{year}{2011}\natexlab{}.
\newblock \showarticletitle{Precomputing Possible Configuration Error Diagnoses}. In \bibinfo{booktitle}{\emph{Proceedings of the 2011 26th IEEE/ACM International Conference on Automated Software Engineering}} \emph{(\bibinfo{series}{ASE '11})}. \bibinfo{publisher}{IEEE Computer Society}, \bibinfo{address}{Washington, DC, USA}, \bibinfo{pages}{193--202}.
\newblock
\showISBNx{978-1-4577-1638-6}
\urldef\tempurl%
\url{https://doi.org/10.1109/ASE.2011.6100053}
\showDOI{\tempurl}


\bibitem[Song and Tilevich(2012)]%
        {Song12}
\bibfield{author}{\bibinfo{person}{Myoungkyu Song} {and} \bibinfo{person}{Eli Tilevich}.} \bibinfo{year}{2012}\natexlab{}.
\newblock \showarticletitle{Metadata Invariants: Checking and Inferring Metadata Coding Conventions}. In \bibinfo{booktitle}{\emph{Proceedings of the 34th International Conference on Software Engineering}} (Zurich, Switzerland) \emph{(\bibinfo{series}{ICSE '12})}. \bibinfo{publisher}{IEEE Press}, \bibinfo{pages}{694--704}.
\newblock
\showISBNx{9781467310673}


\bibitem[Team(2024)]%
        {JUnit5}
\bibfield{author}{\bibinfo{person}{JUnit Team}.} \bibinfo{year}{2024}\natexlab{}.
\newblock \bibinfo{title}{JUnit 5}.
\newblock
\newblock
\urldef\tempurl%
\url{https://junit.org/junit5/}
\showURL{%
\tempurl}
\newblock
\shownote{Accessed: 2024-07-25}.


\bibitem[Weiss et~al\mbox{.}(2017)]%
        {Weiss:2017:TIS}
\bibfield{author}{\bibinfo{person}{Aaron Weiss}, \bibinfo{person}{Arjun Guha}, {and} \bibinfo{person}{Yuriy Brun}.} \bibinfo{year}{2017}\natexlab{}.
\newblock \showarticletitle{Tortoise: Interactive System Configuration Repair}. In \bibinfo{booktitle}{\emph{Proceedings of the 32Nd IEEE/ACM International Conference on Automated Software Engineering}} (Urbana-Champaign, IL, USA) \emph{(\bibinfo{series}{ASE 2017})}. \bibinfo{publisher}{IEEE Press}, \bibinfo{address}{Piscataway, NJ, USA}, \bibinfo{pages}{625--636}.
\newblock
\showISBNx{978-1-5386-2684-9}
\urldef\tempurl%
\url{http://dl.acm.org/citation.cfm?id=3155562.3155641}
\showURL{%
\tempurl}


\bibitem[Wen et~al\mbox{.}(2020)]%
        {Wen20}
\bibfield{author}{\bibinfo{person}{Chengyuan Wen}, \bibinfo{person}{Yaxuan Zhang}, \bibinfo{person}{Xiao He}, {and} \bibinfo{person}{Na Meng}.} \bibinfo{year}{2020}\natexlab{}.
\newblock \showarticletitle{Inferring and Applying Def-Use Like Configuration Couplings in Deployment Descriptors}. In \bibinfo{booktitle}{\emph{2020 35th IEEE/ACM International Conference on Automated Software Engineering (ASE)}}. \bibinfo{pages}{672--683}.
\newblock


\bibitem[Zhang et~al\mbox{.}(2024)]%
        {Zhang2024}
\bibfield{author}{\bibinfo{person}{Huaien Zhang}, \bibinfo{person}{Yu Pei}, \bibinfo{person}{Shuyun Liang}, {and} \bibinfo{person}{Shin~Hwei Tan}.} \bibinfo{year}{2024}\natexlab{}.
\newblock \showarticletitle{Understanding and Detecting Annotation-Induced Faults of Static Analyzers}.
\newblock \bibinfo{journal}{\emph{Proc. ACM Softw. Eng.}} \bibinfo{volume}{1}, \bibinfo{number}{FSE}, Article \bibinfo{articleno}{33} (\bibinfo{date}{jul} \bibinfo{year}{2024}), \bibinfo{numpages}{23}~pages.
\newblock
\urldef\tempurl%
\url{https://doi.org/10.1145/3643759}
\showDOI{\tempurl}


\bibitem[Zhang and Ernst(2013)]%
        {Zhang:2013:ADS}
\bibfield{author}{\bibinfo{person}{Sai Zhang} {and} \bibinfo{person}{Michael~D. Ernst}.} \bibinfo{year}{2013}\natexlab{}.
\newblock \showarticletitle{Automated Diagnosis of Software Configuration Errors}. In \bibinfo{booktitle}{\emph{Proceedings of the 2013 International Conference on Software Engineering}} (San Francisco, CA, USA) \emph{(\bibinfo{series}{ICSE '13})}. \bibinfo{publisher}{IEEE Press}, \bibinfo{address}{Piscataway, NJ, USA}, \bibinfo{pages}{312--321}.
\newblock
\showISBNx{978-1-4673-3076-3}
\urldef\tempurl%
\url{http://dl.acm.org/citation.cfm?id=2486788.2486830}
\showURL{%
\tempurl}

\bibitem[Kabir et~al\mbox{.}(2025)]%
        {zenodo-data}
\bibfield{author}{\bibinfo{person}{Md Mahir Asef Kabir}, \bibinfo{person}{Xiaoyin Wang}, {and} \bibinfo{person}{Na Meng}.}
\bibinfo{year}{2025}\natexlab{}.
\newblock \showarticletitle{MeCheck: A Rule-Based Metadata Bug Detector (Artifact for FSE 2025)}.
\newblock \bibinfo{publisher}{Zenodo}.
\newblock \bibinfo{month}{April}.
\newblock \bibinfo{doi}{10.5281/zenodo.15205192}.
\newblock \bibinfo{howpublished}{\url{https://doi.org/10.5281/zenodo.15205192}}.
\newblock \bibinfo{note}{swhid: swh:1:dir:3cb5cc61c48656a73c90f3aca7cb7d13a5e00d29;origin=https://doi.org/10.5281/zenodo.15205191;visit=swh:1:snp:fc5af7b156e979d17bedfaf050307687a40bbc24;anchor=swh:1:rel:7f61669292af44636411fcbac85fc3ce1867f99f;path=detecting-metadata-bugs}.
\newblock

\end{thebibliography}

\end{document}